\documentclass{aa}
\usepackage{graphicx}
\usepackage{xcolor}
\usepackage{txfonts}
\usepackage{hyperref}
\usepackage[version=3]{mhchem} 
\usepackage[fleqn]{mathtools}  

\makeatletter   
\renewcommand*\aa@pageof{, page \thepage{} of \pageref*{LastPage}}
\makeatother

\newcommand{\hdtwenty}{HD~209458~b}

\newcommand{\Teq}{T_{\textrm{eq}}}
\newcommand{\Teff}{T_{\textrm{eff}}}

\newcommand*\chem[1]{\ensuremath{\mathrm{#1}}}
\newcommand{\kzz}{K_{zz}}
\newcommand{\timescale}[1]{\tau_{\textrm{#1}}}
\newcommand{\degrees}{^\circ}
\newcommand{\figref}[1]{(Fig.~\ref{#1})}                    
\newcommand{\about}{\mathord{\sim}} 

\begin{document}

   \title{Photodissociation and induced chemical asymmetries on ultra-hot gas giants}
  \subtitle{A case study of HCN on WASP-76~b}
  
    \authorrunning{Baeyens et al.}
    \titlerunning{Photochemistry in WASP-76~b}

   \author{Robin Baeyens,\inst{1}
          Jean-Michel D\'esert,\inst{1}
          Annemieke Petrignani, \inst{2}
          Ludmila Carone, \inst{3,4}
          \and
          Aaron David Schneider\inst{5,6}
          }

   \institute{ Anton Pannekoek Institute for Astronomy, University of Amsterdam, Science Park 904, 1098 XH Amsterdam, The Netherlands
              \email{r.l.l.baeyens@uva.nl}
         \and 
         Van `t Hoff Institute for Molecular Sciences, University of Amsterdam, Science Park 904, 1098 XH, Amsterdam, The Netherlands
         \and
         Space Research Institute, Austrian Academy of Sciences, Schmiedlstrasse 6, 8042 Graz, Austria
         \and
         Centre for Exoplanet Science, School of Physics \& Astronomy, University of St Andrews, North Haugh, St Andrews KY169SS, United Kingdom
         \and
         Centre for ExoLife Sciences, Niels Bohr Institute, Øster Voldgade 5, 1350 Copenhagen, Denmark
         \and
         Institute of Astronomy, KU Leuven, Celestĳnenlaan 200D, 3001 Leuven, Belgium
             }

   \date{Received XXX; accepted YYY}

 
  \abstract
   {Recent observations have resulted in the detection of chemical gradients on ultra-hot gas giants. Notwithstanding their high temperature, chemical reactions in ultra-hot atmospheres may occur in disequilibrium, due to vigorous day-night circulation and intense UV radiation from their stellar hosts.}
   {The goal of this work is to explore whether photochemistry is affecting the composition of ultra-hot giant planets, and if it can introduce horizontal chemical gradients. In particular, we focus on hydrogen cyanide (HCN) on WASP-76~b, as it is a photochemically active molecule with a reported detection on only one side of this planet.}
   {We use a pseudo-2D chemical kinetics code to model the chemical composition of WASP-76~b along its equator. Our approach improves on chemical equilibrium models by computing vertical mixing, horizontal advection, and photochemistry. 
   }
   {We find that production of HCN is initiated through thermal and photochemical dissociation of CO and \chem{N_2} on the day side of WASP-76~b. The resulting radicals are subsequently transported to the night side via the equatorial jet stream, where they recombine into different molecules. This process results in an HCN gradient with a maximal abundance on the planet's morning limb. We verified that photochemical dissociation is a necessary condition for this mechanism, as thermal dissociation alone proves insufficient. Other species produced via night-side disequilibrium chemistry are \chem{SO_2} and \chem{S_2}.}
   {Our model acts as a proof of concept for chemical gradients on ultra-hot exoplanets. We demonstrate that even ultra-hot planets can exhibit disequilibrium chemistry and recommend that future studies do not neglect photochemistry in their analyses of ultra-hot planets.}


   \maketitle

\section{Introduction}






The distinct three-dimensional nature of (ultra-)hot Jupiters has been the subject of numerous studies over the past years \citep{Pluriel2023_review}. As these planets are tidally locked, the strong one-sided irradiation they experience shapes their atmospheres, resulting in physical and chemical states that are potentially very different on the day side, night side, evening limb, and morning limb. At the same time, vigourous atmospheric circulation may wash out these differences and result in a more homogeneous atmosphere \citep{Showman2020_review}. Understanding how this complex interplay between radiation, dynamics, and chemistry affects observational signatures is an ongoing research topic in exoplanet science, and an essential one in order to make statements about bulk properties of the atmosphere.

Precise measurements of the ingress and egress of the transit can reveal a non-constant or asymmetric transmission radius \citep{vonParis2016, EspinozaJones2021, GrantWakeford2023_strings}, hence providing constraints on asymmetries between the leading (morning) limb and the trailing (evening) limb. Now for the first time, observations of gaseous exoplanets using the \textit{James Webb Space Telescope} (\textit{JWST}) are reaching the sensitivity required to probe local atmospheric differences directly. 
It is to be expected that limb asymmetries 
in more planets will be studied as the technique matures and \textit{JWST} continues its operation.

Another powerful technique that enables disentangling observational components from different parts of the planetary disk, is high resolution spectroscopy. Since this technique involves tracking the planetary spectral signal as a function of phase during its transit across the star, it becomes possible to separate the beginning and end of transit -- and thus also the leading and trailing limb of the planet. Using this technique, \citet{Ehrenreich2020} and later \citet{KesseliAndSnellen2021} measured a spectral signal of iron in the atmosphere of ultra-hot WASP-76~b (equilibrium temperature $\Teq = 2200~K$) with a radial velocity shift that differed between the morning and evening limbs. Such asymmetry was eventually found to exist for most transition metals on the planet, revealing a chemical limb-to-limb gradient that could be caused by condensation \citep{Kesseli2022_spectralSurvey, Pelletier2023_VO_WASP-76b}. Evidence of chemical and planetary wind asymmetries have likewise been found in another bright ultra-hot exoplanet \citep{Prinoth2022_TiO}. 
Additional studies have established that the interpretation of high-resolution spectra of ultra-hot atmospheres is challenging and requires several detected species, as planetary rotation, wind dynamics, and chemical gradients all impart information on the spectral lines in the form of radial velocity (Doppler) shifts and line broadening \citep{Seidel2021_WASP-76b, WardenierEtal2021mnrasWASP-76bIron, Wardenier2023_WASP-76b, Savel2022_WASP-76b, SavelEtal2023apjAsymmetries, Gandhi2022_WASP-76b}. 

A particularly intriguing result is the detection of hydrogen cyanide (HCN) absorption on only the morning limb of WASP-76~b by \citet{SanchezLopezEtal2022aaWASP76b}. The authors found an HCN signal at 5.2$\sigma$ significance that is red-shifted by $20.8^{+7.8}_{-3.9}$~km/s. In the measured transit geometry, this would constitute gas crossing the morning terminator from the night side to the day side of the planet. The authors hypothesize that, given the difficulty of forming HCN in chemical equilibrium at low carbon-to-oxygen ratio (C/O), this result could signify a strong C/O gradient across the night side, and thus provide further tentative evidence for a condensation scenario. Indeed, \citet{SavelEtal2023apjAsymmetries} have confirmed that a solar elemental composition in chemical equilibrium would not give rise to measurable HCN absorption.

Generally, the assumption of chemical equilibrium in hot exoplanets is reasonable, since disequilibrium chemistry caused by mixing or photodissociations mostly affects cooler planets in which chemical reaction times are comparatively long \citep[e.g.][]{Baxter2021_verticalmixing, Roudier2021}. However, the main production mechanism for HCN in irradiated gaseous exoplanets is through photochemistry. On hot Jupiters, a common HCN formation pathway is via the net reaction \citep{MosesEtal2011apjDisequilibrium}
\begin{equation}
    \chem{CH_4} + \chem{NH_3} \rightarrow \chem{HCN} + 3 \chem{H_2}, \label{eq_ch4_nh3}
\end{equation} which is initiated by the removal of H from \chem{CH_4} and \chem{NH_3} through photodissociation or through photochemically produced atomic hydrogen. This process can yield HCN abundances that are orders of magnitude higher than what is expected from chemical equilibrium. It is clear that the pathway above relies on \chem{CH_4} and \chem{NH_3} being present in sufficient quantities. However, carbon and nitrogen in hot planetary atmospheres ($\Teff \gtrapprox 1500$~K) are mostly locked up in the strong, triple-bonded species \chem{CO} and \chem{N_2}. Hence, unless these species are thermally or photochemically destroyed, HCN formation is suppressed on hotter exoplanets. 

Recently, \textit{JWST} observations of WASP-39~b have revealed the presence of \chem{SO_2}, which represents the first direct evidence of photochemistry in a hot exoplanet atmosphere \citep{Tsai2023_SO2_Nature}. Similar to the mechanism of HCN, \chem{SO_2} is produced if its parent species \chem{H_2S} loses its hydrogen through reactions with atomic H, originating from photolyzed water. Next, reactions with OH radicals -- likewise a product of \chem{H_2O} photolysis -- further oxidize sulfur, resulting in a bulk \chem{SO_2} abundance of 10 - 100~ppm. These findings illustrate the importance of photochemistry in irradiated planets, and suggest that disequilibrium chemistry should be considered for HCN.

Photochemistry may also be a driver of limb asymmetries. Since photochemical dissociations are caused by stellar radiation, mostly in the UV domain, they can only take place on the day side of tidally locked exoplanets. On the other hand, advective processes such as atmospheric dynamics can carry species between the irradiated and obscured hemispheres \citep[e.g.][]{Cooper2006, Steinrueck2019_disequilibrium, Drummond2020, Zamyatina2023, Lee2023_minichemGCM}, such that the chemical composition on the night side is balanced by the advection of photochemical species and the chemical recombination \citep{BaeyensEtal2022mnrasPhotochemistry}. This intrinsic anisotropy of photochemical processes can be a driver of limb asymmetries and chemical gradients, which may be detectable through high-resolution spectroscopy \citep{SavelEtal2023apjAsymmetries}.

Previous studies of giant exoplanet chemistry using \mbox{(pseudo-)2D} chemistry models which incorporate photochemical kinetics, vertical mixing and day-night advection, have concluded that horizontal advection acts as a strong homogenizing factor, quenching the global chemical composition to that of the planetary day side \citep{AgundezEtal2014aaPseudo2D, Venot2020_wasp43b, MosesEtal2021expastPhaseCurvesAriel, BaeyensEtal2021mnrasVerticalHorizontalMixing, BaeyensEtal2022mnrasPhotochemistry, Tsai2021_K2-18b_shallowsurfaces, Tsai2023_SO2_theory}. Although wind systems are excited through irradiation and thus get stronger in hotter planets, the chemical reaction timescales decrease more so, such that cool planets are still more prone to transport-induced chemical disequilibrium. Variations due to photochemistry aside, this balance between dynamical and thermochemical processes results in a nearly homogeneous composition for planets with effective temperatures $\Teff \leq 1400$~K \citep{BaeyensEtal2021mnrasVerticalHorizontalMixing}. Especially at low pressures, such homogeneity is negated when photochemistry is taken into account. Additionally, fast day-to-night-side winds enrich even the night side and morning limb of the planet with photochemically produced species, such as \chem{H}, \chem{HCN}, \chem{C_2H_2} \citep{BaeyensEtal2022mnrasPhotochemistry, KoningsBaeyensDecin2022aaFlares} and \chem{SO_2} \citep{Tsai2023_SO2_theory}. It appears that for these species in particular a photochemical model is preferable to simple chemical equilibrium, motivating our current study.

In this work we investigate whether the recent asymmetric HCN detection by \cite{SanchezLopezEtal2022aaWASP76b} could be driven by intrinsic 3D-ness introduced by photochemistry, if we assume a constant, solar C/O. In addition, we aim to assess if the night sides of ultra-hot giants are in chemical equilibrium. To this end, we construct a pseudo-2D photochemical kinetics model of WASP-76~b (Sect.~\ref{sec_methods}). Our model does not aim to reproduce the observations. Rather, we present abundance maps of several molecular species as a function of longitude, and compare these to a case where photochemical reactions are excluded in order to gain qualitative insight in the role of disequilibrium chemistry on ultra-hot planets (Sect.~\ref{sec_W76}).  Additionally, we present results from a different model which includes sulfur chemistry (Sect.~\ref{sec_sulfur}). Next, we discuss our model results in the context of existing observations (Sect.~\ref{sec_disc_otherobservations}),
and evaluate the implications of WASP-76~b's thermal (Sect.~\ref{sec_disc_thermal}) and wind (Sect.~\ref{sec_disc_winds}) structure. Furthermore, we discuss the impact that ionization may have on the night side of ultra-hot planets (Sect.~\ref{sec_disc_ions}). Finally, we present our conclusions (Sect.~\ref{sec_conclusion}). Additional figures showing distributions of atomic species and a sensitivity test for the vertical mixing are presented in Appendices~\ref{app_atomic} and \ref{app_kzz} respectively.

\section{Photochemical model}
\label{sec_methods}

\begin{figure*}
    \centering
    \includegraphics[width=\textwidth]{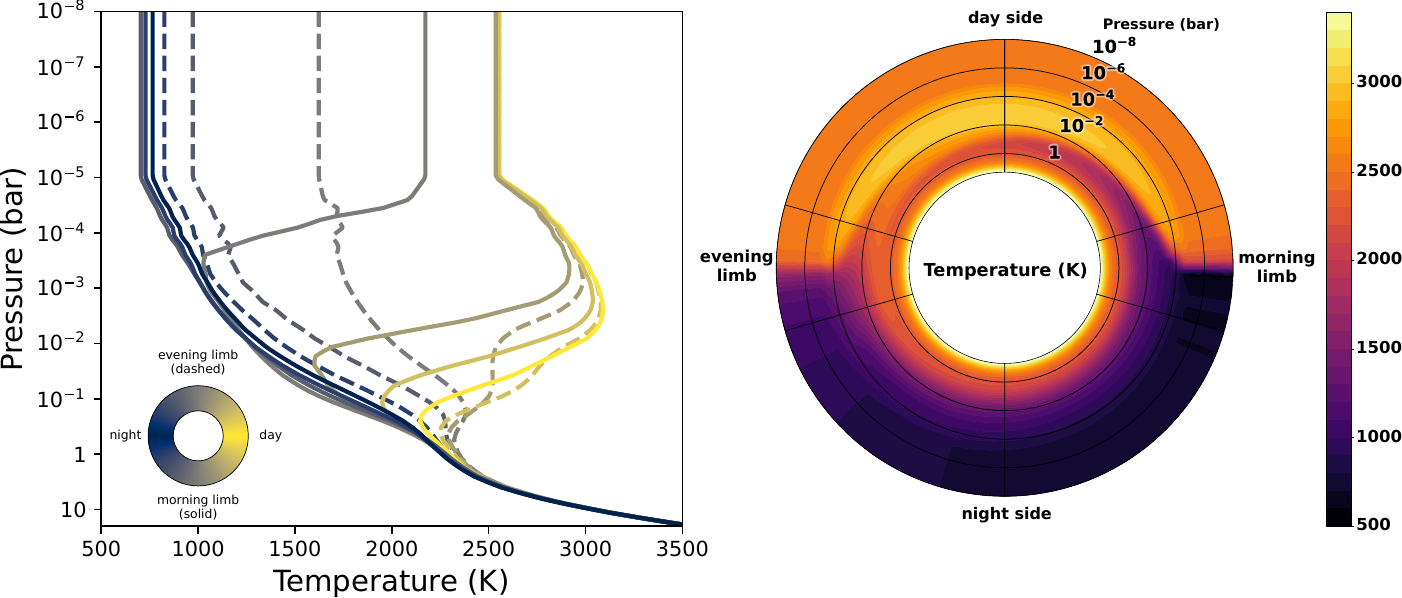}
    \caption{Temperature plots for the WASP-76~b simulation \citep{SchneiderEtal2022aaNoInflation} computed using the \textit{expeRT/MITgcm} \citep{CaroneEtal2020mnrasDeepWinds, SchneiderEtal2022aaExpertMITgcm}. Left: Pressure-temperature profiles plotted at different longitudes on the planetary equator. Solid and dashed lines correspond to longitudes respectively eastward and westward of the substellar point. An isothermal upper atmosphere is assumed for pressures $p < 10^{-5}$~bar. Right: Temperature map as equatorial slice, plotted as a function of longitude and pressure. At the evening and morning limbs, radial lines indicate the opening angle that is probed during transit ($32^\circ$), using the method of \citet{WardenierParmentierLee2022mnrasOpeningAngles}.}
    \label{fig_PT}
\end{figure*}

We construct a photochemical kinetics models for WASP-76~b in order to calculate the HCN abundance, as well as its variation with longitude. For this purpose we employ the pseudo-2D code of \cite{AgundezEtal2014aaPseudo2D}, which allows us to evaluate the effects of vertical mixing, photochemical dissociations, and horizontal advection across the planetary day and night side in a self-consistent way by simplifying the atmospheric winds.

The pseudo-2D configuration is achieved by solving the equation of chemical transport with production and loss terms for each species along a vertical column set at the planetary equator. Then, the column is shifted eastward at a certain pace corresponding to the expected wind speed, to a new longitude on the planetary equator, where the background temperature and irradiation angle are updated. As such, the effect of horizontal advection by the planetary jet stream is incorporated. We stop this procedure once the chemical composition has reached a periodic steady-state.

\subsection{Temperature}

The thermal structure and wind speeds used in our photochemical model are derived from a three-dimensional general circulation model (3D GCM) of WASP-76~b, constructed and presented by \citet{SchneiderEtal2022aaNoInflation}. This model has been computed with the \textit{expeRT/MITgcm} \citep{CaroneEtal2020mnrasDeepWinds, SchneiderEtal2022aaExpertMITgcm} using the correlated-k method to achieve a self-consistent coupling between the primitive hydrodynamic equations and radiative transfer. For further details regarding the setup of the 3D climate model, we refer the reader to \citet{SchneiderEtal2022aaNoInflation}. For post-processing the GCM results and setting up the chemical model, we employ the open-source library \textit{gcm\_toolkit}\footnote{\url{https://github.com/exorad/gcm_toolkit}, \citep{SchneiderBaeyensKiefer2022zenodoGcmtoolkit}}.

For our 2D description of the pressure- and longitude-dependent temperature structure, we have computed the area-weighted averages of the temperature along a $\pm20^\circ$ latitudinal band centred at the planetary equator. Additionally, the temperatures are time-averaged over the final 100 simulation days. Pressure-temperature profiles sampled at different longitudes show a common deep adiabatic gradient at high pressures ($p > 1$~bar) and an increasingly large day-night temperature contrast for lower pressures \figref{fig_PT}. In the case of WASP-76~b, a day-side stratosphere is observed with a strong temperature inversion between 1~bar and 1~mbar, in line with observations of this planet \citep{EdwardsEtal2020ajARES1WASP-76b, FuEtal2021ajWASP-76b, MayKomacekEtal2021ajPhaseCurveWASP-76b, Yan2023_CO_WASP-76b_WASP-18b}. We note that the deep adiabat ($p \gg 1$~bar) in this model has come under scrutiny recently, as longer runtimes would lead to a better convergence with a higher internal temperature \citep{Sainsbury-Martinez2023_SchneiderCarone}. The difficulty of attaining a converged interior in a 3D GCM is a long-standing problem, but it is not considered to be an issue for this study, because of a general decoupling of the interior from the upper layers studied here \citep[see also][]{Sainsbury-Martinez2019, CaroneEtal2020mnrasDeepWinds, Komacek2022_tint}.

In addition to the temperature data from the GCM, which has a vertical upper boundary at $p = 10^{-5}$~bar, we added an isothermal upper atmosphere extension up to $10^{-8}$~bar \figref{fig_PT}. Since hot Jupiter GCMs become unreliable at very low pressures, due to the breakdown of the hydrostatic approximation and interactions with the upper boundary, such a parametrized extension is necessary to investigate the photochemistry of the upper atmosphere. In \citet{BaeyensEtal2022mnrasPhotochemistry} we employed an isothermal upper atmosphere extension as well, and have tested the impact of a hot, day-side thermosphere. We found that that the impact is local, and outside the hot thermosphere there is little effect of the upper atmosphere on the planet's chemistry. We return to this point in Sect.~\ref{sec_disc_thermal}.

Our WASP-76~b setup shows a strong day-night dichotomy in its thermal structure (Fig.~\ref{fig_PT}), as is expected for this class of ultra-hot planets \citep[see e.g.][]{Pluriel2022_day_night_biases, BaeyensEtal2021mnrasVerticalHorizontalMixing,Helling2023_grid}. The minimal and maximal temperature in our simulation ranges from $\about700$~K at the night side near the morning terminator to $\about3000$~K at the substellar point. An important remark is that, adopting the model of \citet{SchneiderEtal2022aaNoInflation}, we did not include the potentially important feedback effect of \chem{H_2} dissociation and recombination. As such, the temperature contrast in our model could be overestimated by several 100~K \citep{Bell2018, Tan2019_andKomacek_ultrahot, Roth2021}. We return to this point in Sect.~\ref{sec_disc_thermal}. 

Finally, we note that our adopted GCM model does not include clouds, although simulations of ultra-hot planets show that even on these high-temperature objects patchy clouds may form on the night side \citep{Komacek2022_patchyclouds, Helling2023_grid}. If these clouds are optically thick, they can affect species detection and asymmetry \citep{Savel2022_WASP-76b, Wardenier2023_WASP-76b}. The presence of night-side clouds may alter the local chemistry in two main ways: through the temperature feedback, and through chemical depletion from the gas phase. Regarding the former, we provide a brief discussion on the night-side temperature uncertainty in Sect.~\ref{sec_disc_thermal}. Regarding the latter, the local carbon-to-oxygen ratio in particular may change through condensation or evaporation of mineral clouds. \citet{Helling2021_cloudtrends} show that cloud-chemistry feedback on ultra-hot Jupiters may cause a slight elevation of C/O~$\approx 0.7$ on the night side as compared to the solar C/O~$\approx0.55$ on the day side. Although this could introduce small quantitative changes in the night-side chemistry, C/O remains below one, avoiding a big shift in chemical regimes. Hence, our use of a cloud-free model is well motivated.

\subsection{Winds and mixing}

\begin{figure}
    \centering
    \includegraphics[width=\columnwidth]{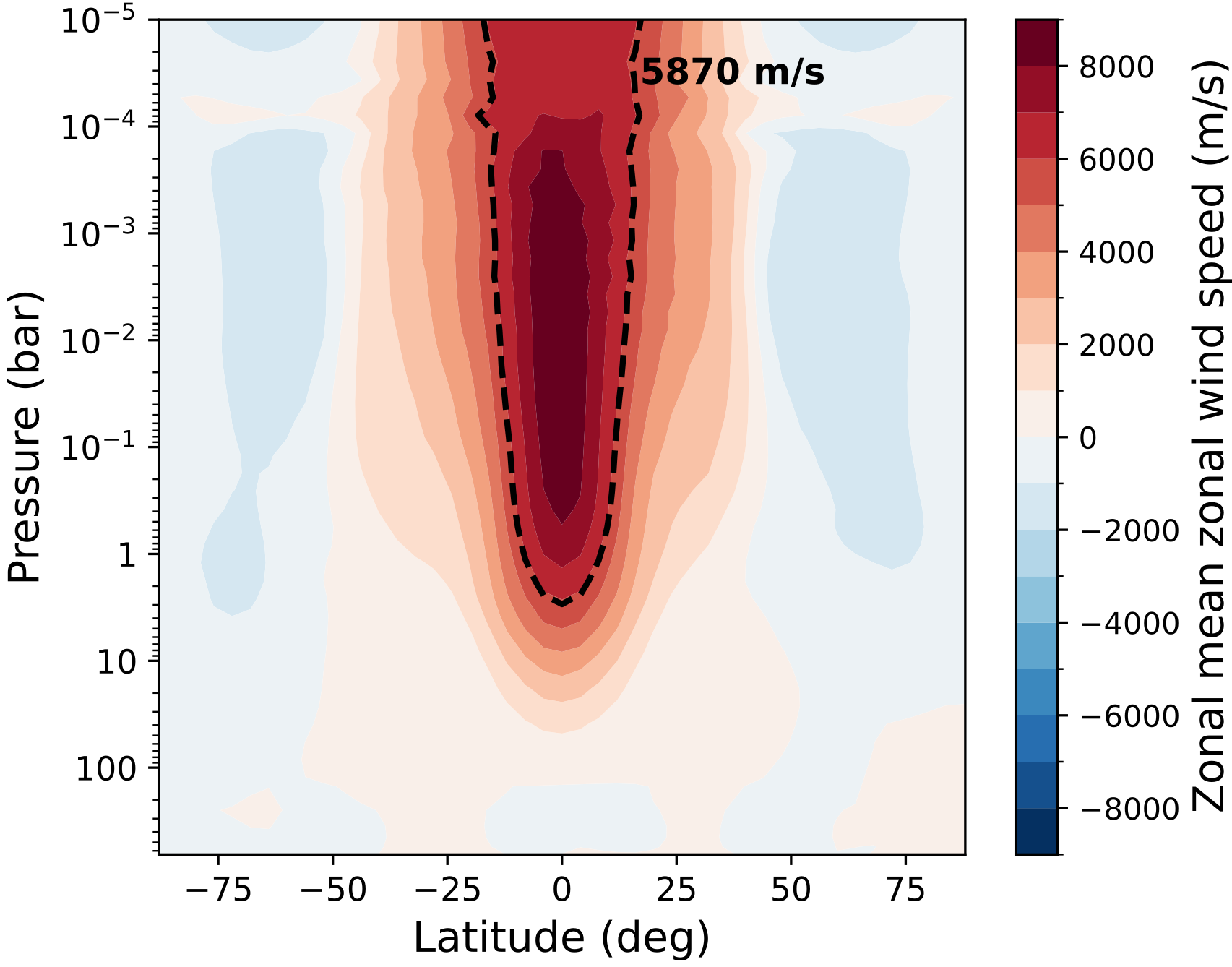}
    \caption{The zonally averaged zonal wind speeds in our 3D climate model of WASP-76~b \citep{SchneiderEtal2022aaNoInflation} show equatorial superrotation throughout the pressure domain. The dashed black line denotes the 5.87~km/s contour that we adopt as the uniform zonal wind speed in our pseudo-2D chemistry model.}
    \label{fig_zonal_wind}
\end{figure}

We derive the horizontal advection speed in our pseudo-2D chemistry code from the mean equatorial wind speed in the GCM. To this end, we compute a weighted average of the zonal wind speed over all longitudes, all latitudes between $\pm20\degrees$, and all pressures above $p<10$~bar. This yields a mean wind jet speed of 5.87~km/s for WASP-76~b.

In the vertical direction, we incorporate vertical mixing through the eddy diffusion coefficient $\kzz$, as is standard practice within the field \citep{ZhangXi2020raaAtmosphericRegimes}. With this parameter, atmospheric mixing by the large-scale overturning motions in hot Jupiter atmospheres is approximated as a diffusion process. While several studies using 3D GCMs have provided numerical and theoretical estimates of $\kzz$ \citep[e.g.][]{ParmentierEtal2013aa3DMixing, CharnayEtal2015apjGJ1214bAnti-Handley, ZhangShowman2018apjIITidallyLocked, KomacekShowmanParmentier2019apjVerticalMixing}, the parameter remains ill-constrained. For our chemistry models, we choose the simple parametrization obtained by \citet{ParmentierEtal2013aa3DMixing} for \hdtwenty{}: $\kzz = 5\cdot 10^8 \left( P_\textrm{bar} \right)^{-0.5}$~cm$^2$\ s$^{-1}$, where $P_\textrm{bar}$ is the pressure in bar. We limit $\kzz$ to a maximum of $10^{11}$~cm$^2$\ s$^{-1}$ in the upper atmosphere.

In the \citet{SchneiderEtal2022aaNoInflation}-climate model used in this paper, the circulation regime of WASP-76~b is dominated by strong equatorial superrotation throughout the range of modelled pressures, the deep interior excepted (Fig.~\ref{fig_zonal_wind}). The zonal winds attain values of several km/s in the equatorial region, with slower retrograde flow in off-equatorial latitudes. As a result, the horizontal flow features little divergence or convergence, and vertical wind speeds are relatively slow ($\sim 10-100$~m/s). 
The consistently fast wind jet at the equator motivates our pseudo-2D approach, where we model horizontal transport as a uniform wind.

Nonetheless, it is likely that the equatorial jet transitions into a regime of thermally direct day-to-night-side flow at low pressures. This transition in GCMs will depend on assumptions of the strength and specific implementation of atmospheric drag. Drag may be used to stabilize the flow near the upper boundary \citep[using a \textit{sponge layer}, ][]{CaroneEtal2020mnrasDeepWinds, Deitrick2020}, or may be motivated by magnetic interactions \citep{Rauscher2013_andMenou_ohmic, Tan2019_andKomacek_ultrahot, Beltz2022_WASP-76b_magneticdrag}. The adoption of strong drag appears to help in matching GCMs to observed phase curves of WASP-76~b \citep{MayKomacekEtal2021ajPhaseCurveWASP-76b}, but high-resolution Doppler shift measurements favour a model with weak atmospheric drag \citep{Savel2022_WASP-76b}. In this work, we used a soft sponge layer (damping only the mean-flow perturbations) and did not assume magnetic drag in our model. We discuss the limitations of our assumption of uniform superrotation in Sect.~\ref{sec_disc_winds}.

\subsection{Chemical network}

For the chemical kinetics, we use the reaction network presented in \citet{Venot2020_network}. This network contains 108 different species, made up of H-, C-, N-, and O-atoms. They are linked together with 1906
reactions: 948 reactions are thermodynamically reversed and 10 are irreversible reactions. We also include 52 photolysis reactions for 35 different species. The chemical network has been experimentally validated, and a comparison with other exoplanet chemistry networks can be found in \citet{Tsai2021}.

To compute the photodissociation rates, we use photoabsorption cross-sections and quantum yields that originate from laboratory measurements
and theoretical calculations \citep{Venot2012, Hebrard2013, Dobrijevic2014}. We do not include temperature-dependent cross-sections. 
As a proxy for the stellar spectrum of WASP-76 (F7 spectral type), we use the same approach as in \citet{BaeyensEtal2022mnrasPhotochemistry} and \citet{KoningsBaeyensDecin2022aaFlares}, which is to make a composite spectrum consisting of a solar-metallicity PHOENIX model \citep{Husser2013} with $\Teff=6500$~K and $\log g = 4.5$ (wavelengths up to 200~nm), an observed UV spectrum of HD~128167 constructed by \citet{Segura2003} (wavelengths between 200 and 115~nm), and a scaled solar spectrum (wavelengths between 115 and 1~nm). The end result is a spectrum representative of a typical F-type star with an XUV flux of $10^{30}$~erg/s. We note that the use of proxied stellar spectra rather than UV observations of the actual exoplanet hosts is not ideal. Hence, obtaining accurate and time-resolved measurements of exoplanet host spectra is an important and ongoing effort \citep[e.g.][]{Youngblood2023_abstract, Behr2023_abstract}. 

The chemical network, photochemistry cross-sections, and stellar spectra used in this work are the same that have been applied in \citet{BaeyensEtal2022mnrasPhotochemistry} and \citet{KoningsBaeyensDecin2022aaFlares}, and additional details may be found there. 

\subsection{Additional model setup details}

As planetary system parameters for WASP-76~b that are used for the chemical model, we have adopted a planetary mass and radius of 0.92~M$_\textrm{Jup}$ and 1.83~R$_\textrm{Jup}$ respectively \citep{WestEtal2016aaWASP-76bWASP-82bWASP-90b}. Furthermore, to compute the stellar irradiation and photochemical rates, we used a semi-major axis value of 0.0330~AU, a proxied F-star spectrum (see above), and a stellar radius of 1.73~R$_\odot$ \citep{WestEtal2016aaWASP-76bWASP-82bWASP-90b}.

The chemical model has been set up with a logarithmically-spaced vertical grid between 20~bar and $10^{-8}$~bar, consisting of 120 layers. Furthermore, in the longitude dimension, we use 90 longitude samples with 4$\degrees$ spacing to sample the changing temperature structure and zenith angle.

As initial condition, we first run a one-dimensional chemical kinetics model, including vertical mixing and photochemistry, which corresponds to the substellar point of the planet. Such starting condition results in an efficient convergence to a periodic steady-state in the pseudo-2D setup \citep{Agundez2012}. For our nominal model, we use a chemical mixture with solar composition, metallicity, and C/O \citep{Asplund2009}. 


\section{Photochemical kinetics of WASP-76~b}
\label{sec_W76}

\begin{figure*}[!p]
    \centering
    \includegraphics[width=\textwidth]{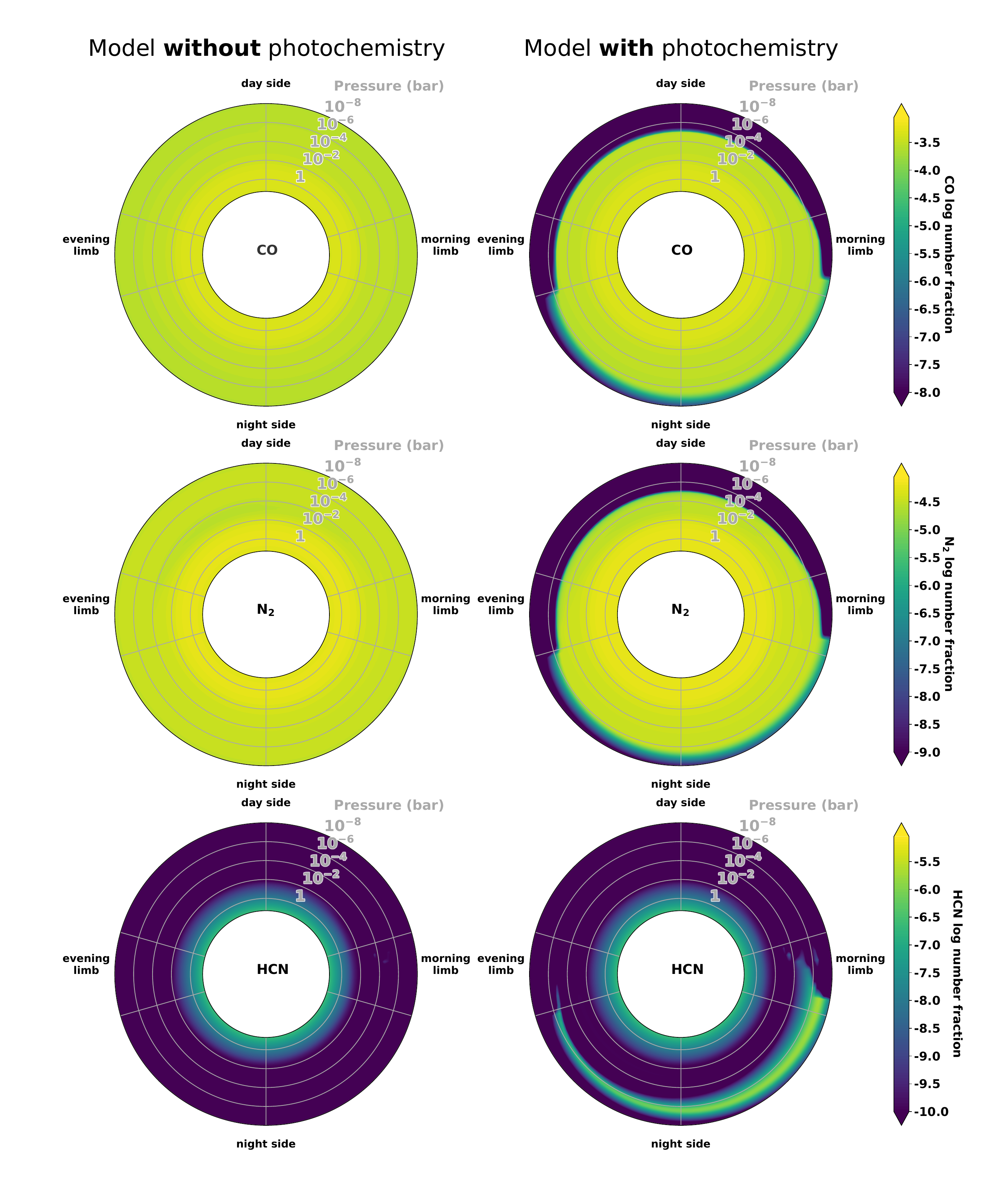}
    \caption{Polar plots of CO, \chem{N_2}, and HCN abundances in WASP-76~b. The left-hand column shows a model that does not incorporate photochemistry, the right-hand column shows the full model with photochemistry. Photodissociation forms a strong additional destruction mechanism for day-side species. HCN formation on the planetary night side is only initiated when photochemistry is taken into account. All plots display an equatorial slice of the planet. At the evening and morning limbs, radial lines indicate the opening angle that is probed during transit ($32^\circ$).}
    \label{fig_chem_CO_N2_HCN}
\end{figure*}

\begin{figure*}[!p]
    \centering
    \includegraphics[width=\textwidth]{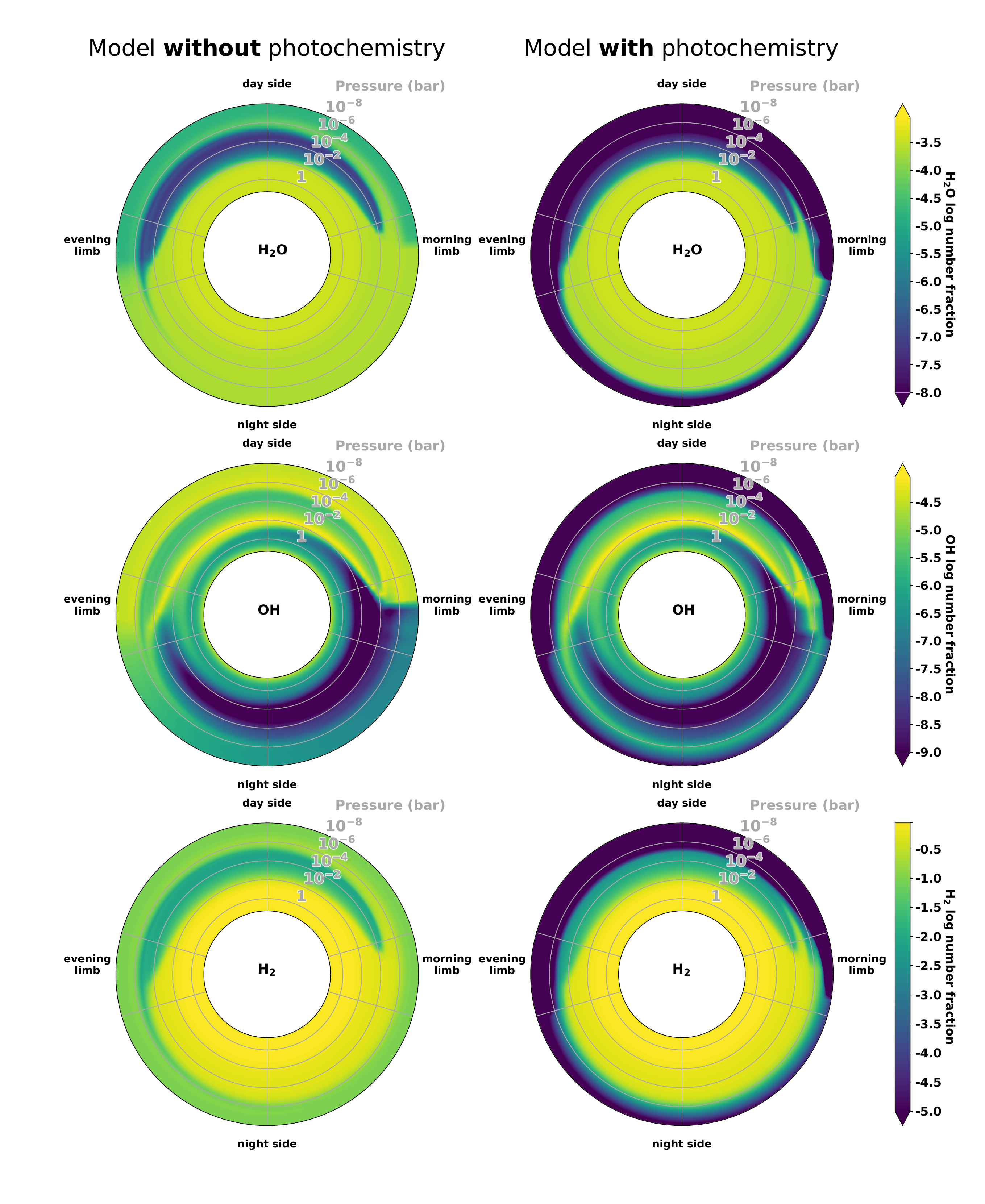}
    \caption{Polar plots of \chem{H_2O}, \chem{OH}, and molecular hydrogen abundances in WASP-76~b. The left-hand column shows a model that does not incorporate photochemistry, the right-hand column shows the full photochemical model. Photodissociation forms a strong additional destruction mechanism for molecular hydrogen and water. All plots display an equatorial slice of the planet. At the evening and morning limbs, radial lines indicate the opening angle that is probed during transit ($32^\circ$).}
    \label{fig_chem_H2O_OH_H}
\end{figure*}

We present an overview of our chemical model as color maps in Figs.~\ref{fig_chem_CO_N2_HCN} (\chem{CO}, \chem{N_2}, \chem{HCN}) and \ref{fig_chem_H2O_OH_H} (\chem{H_2O}, \chem{OH}, \chem{H}). Furthermore, we present atomic abundances in Appendix~\ref{app_atomic}. In general, we find high abundances of CO ($\about10^{-3.5}$), \chem{H_2O} ($\about10^{-3.5}$), and \chem{N_2} ($\about10^{-4.5}$). This result is in line with expectations for a very hot gas planet, as these species are thermodynamically favoured \citep{Lodders2002}. Indeed, other simple, but thermally less stable species like \chem{CH_4} and \chem{NH_3} are virtually absent in our model of WASP-76~b and remain low in abundance ($< 10^{-7}$, not shown here).

\subsection{Thermal dissociation and photodissociation}

Molecular dissociation in the upper atmosphere is apparent in several cases. In order to be able to distinguish between thermal dissociation and photodissociation, we have produced both a fully photochemical model, and a chemical kinetics model where photochemistry is not taken into account (vertical and horizontal mixing are still present). 

For the model without photochemistry (left column of Fig.~\ref{fig_chem_H2O_OH_H}), we find that water and \chem{H_2} start to get thermally dissociated at pressures lower than $p \lesssim 10^{-2}$~bar. This pressure roughly corresponds to the hottest region of the atmospheric day side (Fig.~\ref{fig_PT}), and is in line with theoretical and observational work of thermal dissociation in several ultra-hot giant exoplanets \citep{ParmentierEtal2018aaThermalDissociation}. At the same pressure location, the hydroxyl radical OH reaches a local maximum ($\about10^{-4}$) and becomes more abundant than water. In general, an anti-correlation between OH and \chem{H_2O} can be seen, demonstrating OH to be the product of water dissociation. Between $10^{-2}$ and $10^{-5}$~bar, this transition is thermally driven, as both \chem{H_2O} and \chem{OH} stay close to their equilibrium concentrations, favouring the latter. At pressures lower than $p< 10^{-5}$~bar, photodissociation of both species causes a strong decline. Despite horizontal mixing in our model, only the day-side of the planet shows strong \chem{H_2O} or \chem{H_2} depletion. At the limbs, efficient recombination is taking place. At high altitudes we likewise observe a partial recombination of these species, which is likely a result of our isothermal upper atmosphere assumption.

On the other hand, CO and \chem{N_2} both have very strong triple bonds, making them stable against thermal dissociation. This is evident by their continually high, uniform number fraction in the upper atmosphere of the model without photochemistry (Fig.~\ref{fig_chem_CO_N2_HCN}). Here, however, we observe a striking difference with the fully photochemical model. The model that incorporates photochemistry shows a strong day-side depletion of all molecules, including CO and \chem{N_2}. Concentrations of these species drop by four orders of magnitude at pressures below $p < 10^{-5}$~bar, resulting in an upper atmosphere that is completely atomic in composition. Indeed, dominant molecules are fully dissociated in the upper atmosphere on the day side, following
\begin{align}
    \ce{
        CO + h$\nu$ &-> C + O       \label{eq_diss_co}\\
        N2 + h$\nu$ &-> N + N       \label{eq_diss_n2}\\
        H2O + h$\nu$ &-> H + OH,    \label{eq_diss_h2o}
        }      
\end{align}
with OH itself being photodissociated or participating in the catalytic destruction of \chem{H_2} \citep[see][]{MosesEtal2011apjDisequilibrium}. This chemical transition is clearly seen in abundance maps of the free atoms (Fig.~\ref{fig_app_atomic}), which show a strong day-side enhancement of all atomic species in our network. The difference in atomic carbon between the two cases (with/without photochemistry) is stark, since photodissociation is the key factor causing its release from the abundant CO.

Also on the night side of the planet, upper atmosphere depletion of most main molecules can be observed in the cases where photochemistry is computed (Figs.~\ref{fig_chem_CO_N2_HCN} and \ref{fig_chem_H2O_OH_H}). We do see (partial) recombination of species that have been depleted on the day side, but distinctly less so than in the case without photochemistry. At the lowest pressures in our simulation domain ($p < 10^{-7}$~bar), the night-side hemisphere is still mostly atomic, caused by a slow recombination at low pressures and fast horizontal advection from the day side. Thus, it is clear that photodissociation plays a crucial role in the transition from a molecular to an atomic atmosphere.

\subsection{Night-side HCN}
\label{sec_results_HCN}

Comparing the HCN abundance in the models with and without photochemistry, an interesting picture arises (Fig.~\ref{fig_chem_CO_N2_HCN}). If no photochemistry is used, the model yields almost no HCN. This is not surprising in itself, since thermochemical equilibrium does not favour HCN production at high temperatures and solar abundances ($\text{C/O} = 0.55$). But in the photochemical model, a distinct night-side region between $10^{-4}$~bar and $10^{-7}$~bar does show an elevated HCN number fraction, reaching a maximal concentration of $10^{-6}$ at the morning limb of WASP-76~b. Thus, the presence or absence of night-side HCN hinges on whether photochemistry is taken into consideration in the model.

The HCN enhancement seen in Fig.~\ref{fig_chem_CO_N2_HCN} is asymmetric and it increases with longitude along the night side. This crescent is the result of HCN forming through the recombination of species that have been dissociated on the day-side hemisphere, and subsequently advected to the night side by the equatorial wind flow. Following the prograde wind advection, efficient HCN production starts taking place after crossing the evening limb, as the temperature drops below $\about1000$~K, and continues up to the morning terminator as the thermal background keeps getting cooler. The main pathway of HCN formation on the planetary night side is through recombination of free hydrogen, carbon, and nitrogen atoms originating from the day side. These free atoms recombine on the night side via the following scheme\footnote{These and following reaction analyses are based on the \textit{VULCAN C-H-N-O-S} network \citep{Tsai2021}. See also Sect.~\ref{sec_sulfur}.}:
\begin{align}
    \ce{
        N + O  &->[( + M)] NO       \label{eq_step1}    \\
        C + NO &->[    ] CN + O     \label{eq_step2}    \\
        CN + H &->[( + M)] HCN      \label{eq_step3}
        }
        \shortintertext{\leftline{\rule{6cm}{0.6pt}}} 
    \text{Net:}\quad 
    \ce{
        H + C + N  &-> HCN.         \label{eq_net}
        }      
\end{align}
Here, \chem{M} signifies any third body. In this reaction scheme, the slowest, rate-limiting step is (\ref{eq_step2}), which occurs at a reaction rate of 6~s$^{-1}$cm$^{-3}$ at 1~$\mu$bar. This yields a chemical HCN formation timescale of $\sim$10$^6$~s. Of comparable importance to the formation of the CN radical, and hence HCN, is the reaction \ce{N + CS -> CN + S}, which occurs at a similar reaction rate as (\ref{eq_step2}). On the other hand, the reaction \ce{NO + CN -> CO + N2}, which removes both species driving the HCN formation scheme, is 100 times slower. HCN has destruction pathways through reactions with H and O. In the former case, it is efficiently recycled back to HCN, but in the latter case \ce{HCN + O -> CO + NH} results in locking up the carbon in CO. As the morning terminator is crossed, HCN is again removed rapidly on the day side in favour of CO and \chem{N_2} ($p > 10^{-5}$~bar) or an atomic gas mixture ($p > 10^{-5}$~bar).

Combined with the photodissociation reactions on the day side (\ref{eq_diss_co})-(\ref{eq_diss_h2o}), the net reaction for HCN formation in ultra-hot Jupiters reads
\begin{align}
    \ce{ CO + N2 + H -> HCN + N + O
        }      
\end{align} This reaction is the equivalent of (\ref{eq_ch4_nh3}) in high-temperature, high-radiation environments such as ultra-hot Jupiters.
It is telling that photochemistry is necessary to produce night-side HCN. Without photochemistry, there is no influx of dissociated parent species out of which HCN can be assembled. In the hot atmosphere of WASP-76~b, \chem{CH_4} and \chem{NH_3} are sparse, so freed carbon and nitrogen from CO and \chem{N_2} are required. Only in the photochemical model is there sufficient molecular dissociation of \chem{CO} and \chem{N_2}, evidenced by the upper atmosphere depletions on the day side (Fig.~\ref{fig_chem_CO_N2_HCN}), to initiate HCN formation.


\subsection{Effect of the horizontal wind speed}
\label{sec_jet_speed}

Since the night-side HCN abundance seen in Fig.~\ref{fig_chem_CO_N2_HCN} is increasing toward the morning terminator, it is also increasing with time in our pseudo-2D setup, as the gas is advected along the wind jet. As such, we should expect to see an effect of the jet speed on the chemistry at the night side. In particular, we expect to see a higher HCN abundance at lower wind jet speeds, as there is more time available for HCN to form from the dissociated products.

\begin{figure}
    \centering
    \includegraphics[width=\columnwidth]{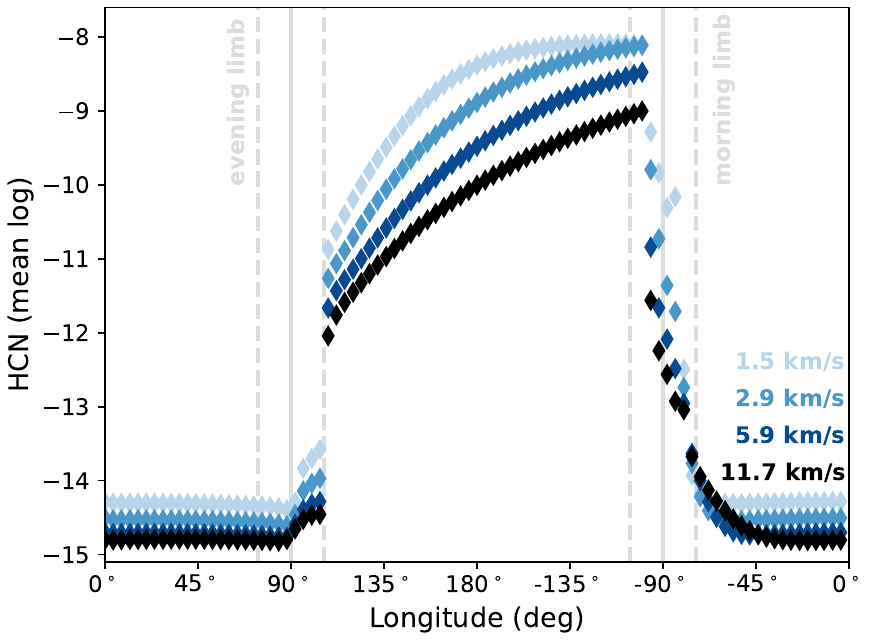}
    \caption{The arithmetic mean of the HCN log number fraction (for $p < 100$~mbar) as a function of longitude is plotted for models with four different jet wind speeds: 1.5~km/s, 2.9~km/s, 5.9~km/s (nominal model), and 11.7~km/s. A clear effect on the night-side HCN abundance is observed. The morning and evening limbs are indicated, including the opening angles using dashed grey lines.}
    \label{fig_vary_wind}
\end{figure}

In order to test this hypothesis, we have ran additional models of WASP-76~b with $0.25 \times$, $0.5 \times$, and $2 \times$ the nominal jet wind speed. In doing so, we effectively vary the horizontal advection timescale from $\sim$10$^6$~s (slowest jet speed) to $\sim$10$^5$~s (fastest jet speed). A clear influence of the wind speed on the night-side HCN abundance is observed (Fig.~\ref{fig_vary_wind}). Specifically, a lower wind speed yields consistently more HCN on the planetary night side. Additionally, the model with the slowest wind jet (1.5~km/s) displays a \textit{plateau} in the HCN abundance. Indeed, the amount of HCN in this model reaches a maximum near the antistellar point, and maintains a fairly constant value during the remainder of its night side crossing. Conversely, all other models show continually increasing HCN fractions with a maximal value attained near the morning limb ($\about$-100$^\circ$). Here, the 2.9~km/s-model is seen to match up again with the slowest model.

We interpret this finding as a balance between the chemical formation timescale and horizontal advection timescale. At slow jet speed, the chemical timescale of HCN formation ($\sim$10$^6$~s, see Sect.~\ref{sec_results_HCN}) is comparable to the advection timescale. Consequently, a sufficiently long time is spent on the cool night side for HCN to form and build up, before the hot temperatures of the day side are reached and HCN is again destroyed. As such, the mean HCN abundance is higher at lower jet speeds, and even saturates at the lowest speed we have tested.



\section{Sulfur photochemistry}
\label{sec_sulfur}

In addition to our nominal model, computed with the \citet{Venot2020_network} chemical network, we present a pseudo-2D photochemical model that makes use of the \textit{VULCAN} C-H-N-O-S network \citep{Tsai2021}. This chemical network contains sulfur-bearing species and has recently been used to model the \chem{SO_2} abundance on the warm gas giant WASP-39~b \citep{Tsai2023_SO2_Nature, Tsai2023_SO2_theory}. On this planet, \chem{SO_2} was detected through \textit{JWST} transmission spectroscopy \citep{Alderson2023_WASP-39b_NIRspec, Rustamkulov2023_WASP-39b_PRISM}, and its abundance has been explained as a result of photochemical reactions transforming the abundant \chem{H_2S} into \chem{SO_2} \citep{Tsai2023_SO2_Nature}. Except for substituting the species and reaction network in our photochemical code, our sulfur-model has been set up in the same way as our nominal model of WASP-76~b.

\begin{figure}
    \centering
    \includegraphics[width=\columnwidth]{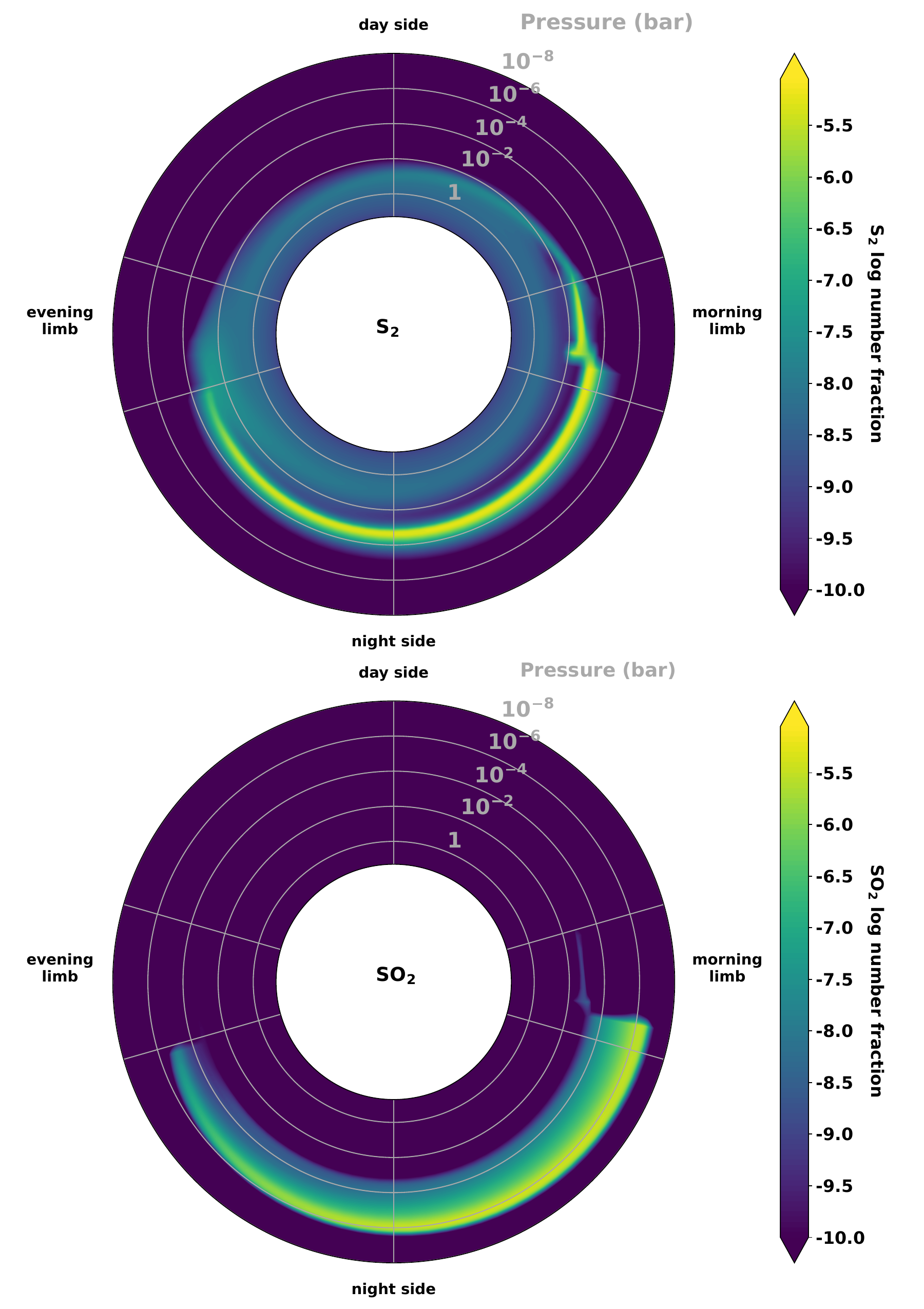}
    \caption{Polar plots of the \chem{S_2} and \chem{SO_2} abundances in WASP-76 b, computed using the C-H-N-O-S chemical network of \citet{Tsai2021} in our pseudo-2D kinetics code. An equatorial slice of the planet is displayed, with radial lines at the limbs indicating the opening angle (32$^\circ$). Enhancements similar to HCN (Fig.~\ref{fig_chem_CO_N2_HCN}) can be seen on the night side.}
    \label{fig_chem_S2_SO2}
\end{figure}

Upon inspecting the abundances of the sulfur-species \chem{S_2} and \chem{SO_2}, we find a similar crescent-shaped enhancement on the night side of WASP-76~b as we found for HCN (Fig.~\ref{fig_chem_S2_SO2}). Both \chem{S_2} and \chem{SO_2} are entirely absent from the day-side hemisphere, where they are destroyed through thermal dissociation and photolysis, but they can survive on the cooler night side, as sulfurous building blocks are supplied via the atmospheric circulation. A maximal abundance of $\about10^{-5}$ is attained at the morning limb of the planet in both cases. This result is reminiscent of the recent two-dimensional WASP-39~b model presented by \citet{Tsai2023_SO2_theory}. There, the authors also found a buildup of \chem{SO_2} on the planet's night hemisphere, and explained this phenomenon as an advection of H and OH radicals from day-side photolysis, which react with \chem{H_2S} to drive the production of \chem{SO_2} on the night side. It appears that a similar phenomenon can take place on WASP-76~b, despite it being a much hotter planet than WASP-39~b. However, in our case, the mechanism occurs via the faster third-body recombination with atomic oxygen:
\begin{align}
    \ce{
        S + O  &->[( + M)] SO       \label{eq_step1_sulfur}    \\
        SO + O &->[( + M)] SO2      \label{eq_step2_sulfur} 
        }
        \shortintertext{\leftline{\rule{6cm}{0.6pt}}} 
    \text{Net:}\quad 
    \ce{
        S + O + O  &-> SO2.           \label{eq_net_sulfur}
        }      
\end{align} Here, the rate-limiting step is the first one, and it is up to 10$^6$ times faster than the combination \ce{S + OH} that underlies the \chem{SO_2} formation on WASP-39~b. 
There are no efficient destruction pathways for \chem{SO_2}, since reactions such as \ce{SO2 + H -> SO + OH} and \ce{SO2 + C -> SO + CO} allow for efficient recycling via (\ref{eq_step2_sulfur}), as long as free oxygen is available. As such, it can build up quickly on the night side.

It is important to note, however, that in our model of the ultra-hot WASP-76~b, photochemistry is not a requirement for the production of night-side \chem{SO_2}. This becomes evident when looking at the distributions of sulfur-bearing species in a model without photochemistry (not shown here). We found that the distributions are virtually unchanged from those shown in the photochemical kinetics model (Fig.~\ref{fig_chem_S2_SO2}). Indeed, thermal dissociation of water on WASP-76~b's day side alone provides sufficient H and OH radicals to break up \chem{H_2S} and free up the necessary sulfur for night-side \chem{SO_2} production.
Hence, unlike for cooler planets like WASP-39~b \citep{Tsai2023_SO2_Nature, Tsai2023_SO2_theory}, the presence of \chem{SO_2} on the night side of our WASP-76~b model does not indicate a photochemical production mechanism. 

\chem{S_2} attains equally high abundances as \chem{SO_2} in our model, but at higher pressures ($\about 1$~mbar). Like \chem{SO_2}, it forms directly via third-body combination of the atoms: \ce{S + S ->[( + M)] } S2. A high \chem{S_2} content is not unexpected in hot gas giants \citep{Zahnle2009_sulfur, Hobbs2021}. In the VULCAN network, \chem{S_2} can continue reacting with elemental sulfur molecules, forming allotropes up to \chem{S_8}. Eventually, this reaction path could lead to condensation into sulfur hazes and soots \citep{Zahnle2009_photo, Zahnle2016, Gao2017_sulfurhaze, Tsai2023_HD80606}. Besides S and \chem{S_2}, however, there are no such species present in non-negligible amounts. Hence, WASP-76~b's night side is -- like WASP-39~b \citep{Tsai2023_SO2_theory} -- too hot for sulfur polymerization or condensation to occur.

\section{Discussion}
\label{sec_discussion}

\subsection{Observational context}
\label{sec_disc_otherobservations}

One of our primary motivations for the present study is to qualitatively explain the intriguing detection of HCN absorption on only the morning limb of WASP-76~b \citep{SanchezLopezEtal2022aaWASP76b}. The nominal model presented in this work represents a proof of concept for this detection, as it naturally explains 1) how HCN may be formed in an atmosphere as hot as WASP-76~b; and 2)
why the absorption is only present on the morning limb of the planet. Nevertheless, several discrepancies between our model and the reported detection remain. Firstly, the modelled HCN distribution forms at very high altitudes, and efficient zonal advection mostly keeps it confined near $p \sim 10^{-6}$~bar (Fig.~\ref{fig_chem_CO_N2_HCN}). Although high-resolution spectroscopy is sensitive to individual lines, and thus tends to probe lower pressures than typical transmission spectroscopy, it is still questionable whether the HCN distribution in our model would cause a detectable signal. 
Stronger vertical mixing would result in the same HCN peak abundance near $\sim$10$^{-6}$~bar, but a higher integrated column density (see Appendix~\ref{app_kzz}). Such a scenario of higher vertical mixing is not unlikely, given the expected trend of higher wind speeds for hotter planetary atmospheres and the potential for a vertical advection \textit{chimney} caused by converging winds on the night side \citep{ParmentierEtal2013aa3DMixing, ZhangShowman2018apjIITidallyLocked, KomacekShowmanParmentier2019apjVerticalMixing}. 
Besides the vertical mixing efficiency, the HCN abundance also varies as a function of wind speed (Sect.~\ref{sec_jet_speed}) as well as night-side temperature (see Sect.~\ref{sec_disc_thermal} below). Thus, further work including a direct comparison to the observational data as a function of key parameters would be necessary to conclude on the origin of the HCN signal. We remark that, besides disequilibrium chemistry, another possibility is that the HCN detection is a false positive, considering the relatively low significance and peculiar radial velocity \citep[see also][]{Zhang2020_Chachan_Kempton_PLATON2_HD18, SavelEtal2023apjAsymmetries}.

\citet{Landman2021_OH_WASP-76b} have reported a detection of the hydroxyl radical (OH) on WASP-76~b at a significance of 6.1, likely originating from the evening terminator. As such, it is a strong indicator of water dissociation, which we also find in our nominal model in a fairly symmetric distribution (Fig.~\ref{fig_chem_H2O_OH_H}). Photochemistry in our model is destroying OH in the upper atmosphere ($p < 10^{-6}$~bar), but thermal dissociation of water near $10^{-2}$~bar is providing a high OH abundance regardless of whether photochemistry is incorporated. The temperature constraints formulated in \citet{Landman2021_OH_WASP-76b} (2700~K-3700~K) are consistent with the GCM used in this work.

\citet{Kesseli2022_spectralSurvey} and recently \citet{Pelletier2023_VO_WASP-76b} have detected a multitude of refractory species on WASP-76~b using high-resolution spectroscopy, among which VO and several ions (e.g.~\chem{Ca^+}, \chem{Fe^+}, and \chem{Sr^+}). Having several simultaneous detections like this provides valuable information on the temperature and ionization state of the atmosphere. Approximately, the molecular/atomic hydrogen transition and oxygen abundance in our model appears to agree with the findings of \citet{Pelletier2023_VO_WASP-76b}, although we do not employ a hot thermosphere of $>3000$~K like they retrieve (see their Extended Data Table~6). Although from a modelling point of view, the location of a molecular-atomic transition may be influenced either by thermal or by photochemical dissociation processes, chemical-radiative feedback in a self-consistent atmosphere will lead to a hot thermosphere that is sustained by both processes (see also following Sect.~\ref{sec_disc_thermal}). 

Another point to note is that our inclusion of photochemical dissociation results in the destruction of high altitude CO and \chem{N_2}. In particular, comparing our photochemical model with a model that only computes thermal dissociation (equilibrium chemistry), we find that the abundances of CO and \chem{N_2} drop by orders of magnitude at $p < 10^{-5}$~bar (Fig.~\ref{fig_dissociation}). Recent high-resolution studies make use of the stability of the CO molecule on WASP-76~b to help break degeneracies and constrain chemical asymmetries on the planet \citep{SavelEtal2023apjAsymmetries, Wardenier2023_WASP-76b}. In particular, the assumption of a constant CO mixing ratio helps to constrain the scale height, allowing to diagnose chemical changes in different species. However, our results show that CO can only be assumed constant at $p > 10^{-5}$~bar (Fig.~\ref{fig_dissociation}), and that it can itself display morning-evening asymmetry, albeit at high altitudes (Fig.~\ref{fig_chem_CO_N2_HCN}). Figure~7 of \citet{Wardenier2023_WASP-76b} shows that a sizeable fraction of the CO line originates from pressures lower than $p < 10^{-5}$~bar. Hence, the assumption of constant CO may not hold true, potentially biasing results. Thus, we caution against viewing the CO abundance as uniform and constant, and recommend to include photochemical processes when dealing with lines that are shaped by the upper atmosphere.

Since we find a large influence of photolysis in the transition from a molecular to atomic atmospheric composition in WASP-76~b, we might assume that photo-ionization plays a similar role in relation to thermal ionization \citep[cfr.][]{Barth2021, Helling2021_HAT-P-7b3_ionization, Helling2023_grid}. As such, we encourage future studies to include ion kinetic chemistry in their analysis. We present a discussion of the implications of ion chemistry on our results in Sect.~\ref{sec_disc_ions}.




\subsection{Chemical feedback on the thermal structure}
\label{sec_disc_thermal}

In our photochemical model of WASP-76~b, we have post-processed a 3D GCM, which has led to two main assumptions regarding the thermal structure. The first is that of an isothermal upper atmosphere extension below $p < 10^{-5}$~bar. The second is that no iteration has taken place between the chemical and thermal computations, so the displacement of heat from molecular dissociation/recombination is not taken into account. We discuss both assumptions and their implications for night-side photochemistry.

\subsubsection{Isothermal upper atmosphere}

Given the need for a higher upper boundary in our photochemical model than what can reasonably be achieved with 3D climate models, we have adopted an isothermal extrapolation of existing pressure-temperature profiles between $10^{-5}$~bar and $10^{-8}$~bar. However, one-dimensional theoretical studies \citep[e.g.][]{GarciaMunoz2007, Koskinen2013, Caldiroli2021_ATES} predict that radiative heating by XUV irradiation can lead to temperatures in the upper atmospheric layers ($p< 10^{-6}$~bar) of $\about10\,000$~K. In addition, detections of alkali lines in the upper atmosphere of WASP-76~b have led to temperature constraints, yielding several  thousand kelvin \citep{Seidel2021_WASP-76b, Kawauchi2022_WASP-76b_alkali}. Such results hint at the existence of a hot thermosphere, but at the same time they are not in clear conflict with the isothermal extension of the hot day side assumed here (see Fig.~\ref{fig_PT}).

\begin{figure}
    \centering
    \includegraphics[width=\columnwidth]{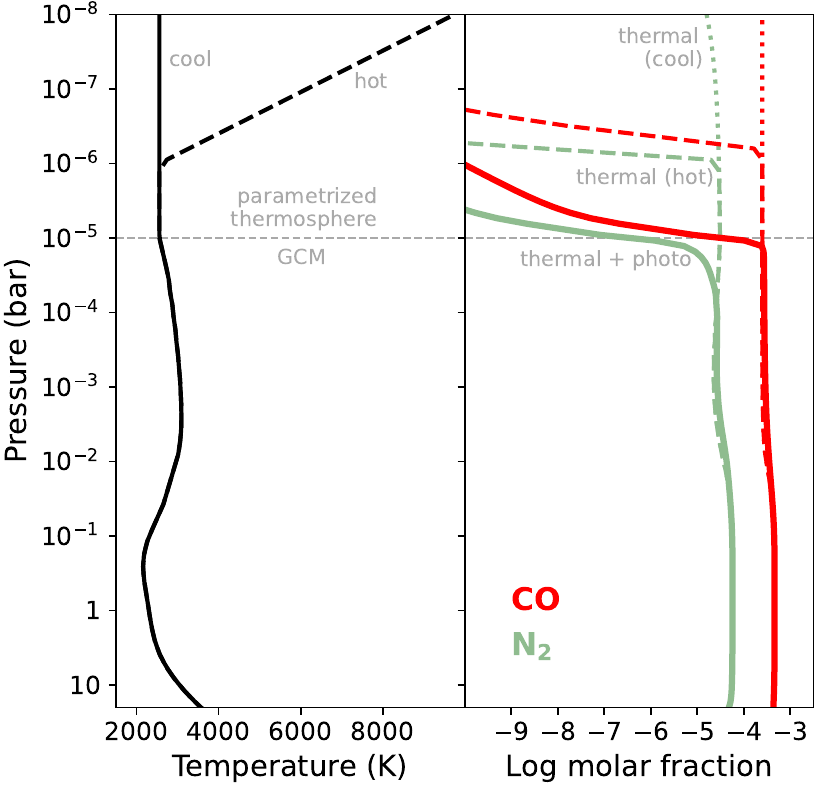}
    \caption{Temperature and abundance profiles of CO and \chem{N_2} at the substellar point of WASP-76~b. Chemical equilibrium models with a cool, isothermal upper atmosphere (\textit{dotted}) show no molecular dissociation, while those with a hot thermosphere (\textit{dashed}) display thermal dissociation at temperatures above $\sim3000$~K. Taking into account photochemistry (\textit{full}) shifts the atomic/molecular transition to higher pressures. The horizontal grey line denotes the region where we adopted a parametrized upper atmosphere.}
    \label{fig_dissociation}
\end{figure}

The biggest impact of a hot thermosphere of several thousand kelvin is the onset of molecular dissociation. At such high temperatures and low densities, even thermally stable species such as CO and \chem{N_2} return to their atomic constituents.
In a previous study, we have tested the impact of the thermospheric temperature on photochemical kinetics, and found only small effects \citep{BaeyensEtal2022mnrasPhotochemistry}. Only where the temperature was changed, local effects on the chemical composition were found. 
In Sect.~\ref{sec_W76} we showed that HCN night-side chemistry relies on the influx of atomic C and N from the day side and argued that photochemistry is needed for the photolysis of CO and \chem{N_2}. We compare thermal dissociation and photodissociation for a planetary model with and without a hot thermosphere. To this end, the thermosphere is parametrized by interpolating the temperature to a maximal value of 10~000~K for pressures between $10^{-6}\text{ bar} > p > 10^{-8}$~bar. We clearly see the importance of photochemistry in pushing the atomic-molecular boundary to higher pressures (Fig.~\ref{fig_dissociation}). In the photochemical kinetics case, however, the lack of radiative cooling by molecules will likely increase the temperature and result in the formation of a hot thermosphere at higher pressures. As such, both effects are very much connected.

Since the influx of atomic building blocks appears safeguarded, either through photodissociation or through thermal dissociation in a hot thermosphere, the question of whether HCN and \chem{SO_2} can form on the night side depends primarily on the night side temperature.
Since the night-side is shielded from direct irradiation, we expect that a thermosphere does not form there and the upper atmosphere does not deviate much from an isotherm. Again, radiative feedback and the presence of molecular coolants on the night side will influence this picture. However, we find that horizontal advection timescales are sufficiently long to allow for (partial) recombination into molecules at $p > 10^{-7}$~bar (Figs.~\ref{fig_chem_CO_N2_HCN} and \ref{fig_chem_H2O_OH_H}). As such, we believe that our ad-hoc assumption of the upper atmospheric temperature does not significantly influences the night-side photochemistry seen in our WASP-76~b model, at least not up to $10^{-7}$~bar. Future 2D or 3D modelling incorporating radiative and chemical feedback would be needed to provide better predictions of the upper atmospheres on the night sides of ultra-hot planets.

\subsubsection{Molecular dissociation/recombination}

Previous studies have shown that the self-consistent internal coupling between chemistry and temperature generally does not affect the thermal structure of the planet much compared to post-processed chemistry; only in specific cases involving \chem{CH_4} quenching are changes of at most 100~K or roughly 10~\% expected \citep{Drummond2016, Drummond2020, Steinrueck2019_disequilibrium, Zamyatina2023}. However, the atmosphere of WASP-76~b is so hot that molecular hydrogen may be dissociated, and subsequently transported to the night side, where latent heat from hydrogen recombination is released \citep{Bell2018, Komacek2018_andTan_researchnote_ultrahot}. 

\citet{Roth2021} have investigated the impact of hydrogen heat capacity and self-consistent chemistry in a pseudo-2D model framework. For a planet with equilibrium temperature $\Teq = 2600$~K, the authors saw day-side cooling by up to 400~K and heating near the evening terminator by up to 800~K. However, in planets with lower equilibrium temperature, such as WASP-76~b ($\Teq = 2200$~K), the impact of hydrogen dissociation and recombination should be lower, as demonstrated by \citet{Tan2019_andKomacek_ultrahot}. Their Fig.~11 shows that the day-night temperature contrast at 2200~K should be comparable to a case where dissociation is neglected. Only at higher equilibrium temperatures the day-night contrast is markedly different than what it would be if hydrogen dissociation and recombination were ignored.

As a reference, we compare our thermal structure with the model of \citet{MayKomacekEtal2021ajPhaseCurveWASP-76b}. The authors of this study likewise computed a 3D GCM for WASP-76~b. In their model, the heat redistribution caused by hydrogen dissociation/recombination is accounted for. However, they adopted a radiative treatment (double-gray) that is simpler than the correlated-k method used in our nominal model of \citet{SchneiderEtal2022aaNoInflation}. Upon comparison, we find good qualitative agreement with the model of \citet{MayKomacekEtal2021ajPhaseCurveWASP-76b}, but with some quantitative differences. The main discrepancy appears to be the night-side temperature, which is $\about750$~K in our case, but not lower than 1000~K in the model of \citet{MayKomacekEtal2021ajPhaseCurveWASP-76b}. It is possible that this temperature difference originates in hydrogen latent heat, and hence, we are underestimating the night-side temperature in our model. Regarding the day-side temperature, on the other hand, we see good agreement with the \citet{MayKomacekEtal2021ajPhaseCurveWASP-76b}-GCM. 

\begin{figure}
    \centering
    \includegraphics[width=\columnwidth]{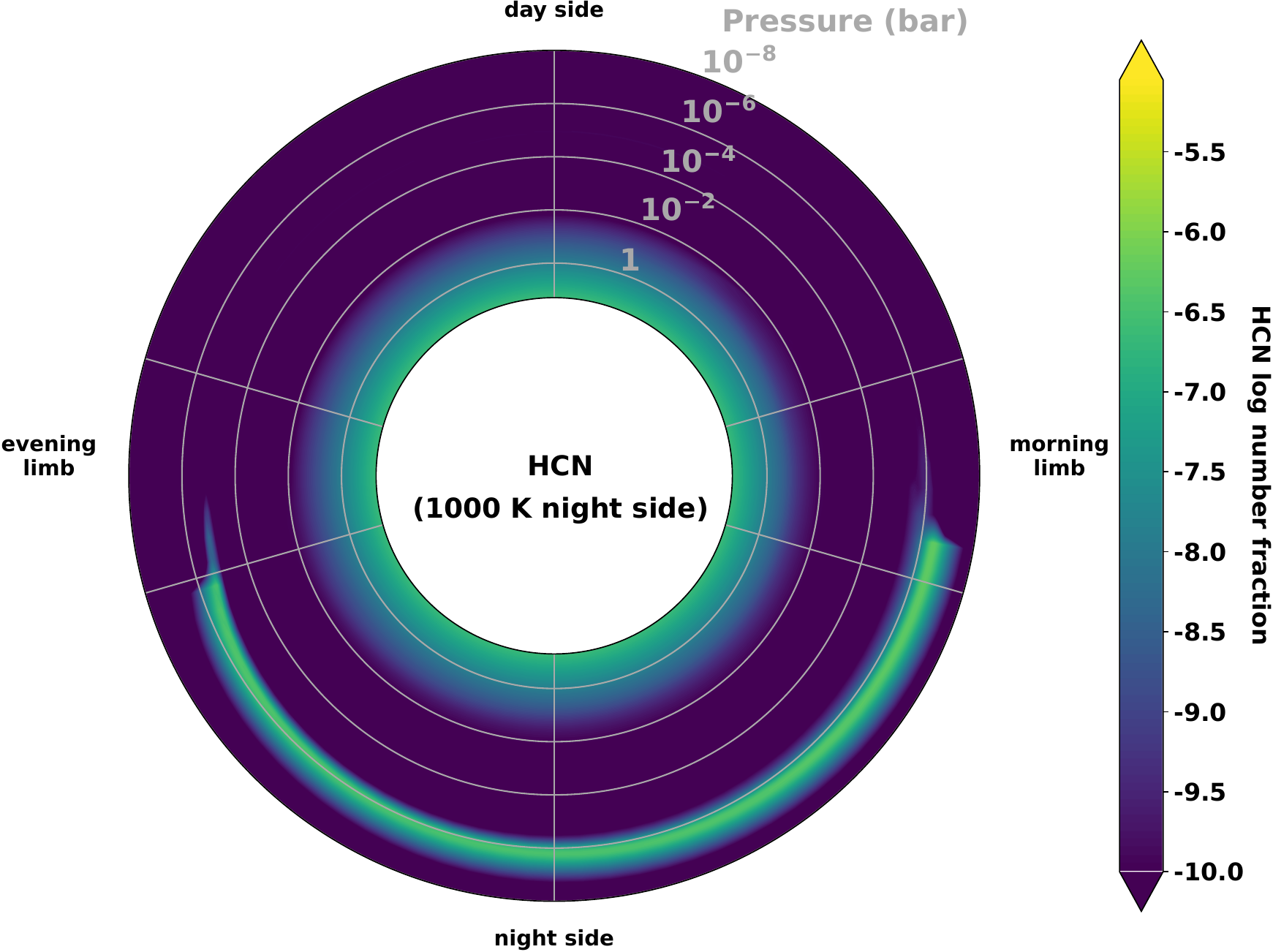}
    \caption{Polar plot of the HCN abundance in a toy model mimicking hydrogen dissociation and recombination. The minimal temperature in this model is set at 1000~K.}
    \label{fig_1000K}
\end{figure}

Since modelling the atmospheric temperature and chemistry of WASP-76~b including full latent heat transport of dissociating and recombining molecules is out of the scope of this study, we instead perform a toy model with increased night-side temperature. Specifically, we set the night-side temperature in the 3D GCM to a minimum of 1000~K and repeat the chemical model, in order to assess the potential effect on the night-side photochemistry. We find a qualitatively similar, crescent-shaped night-side distribution of HCN, including a maximum near the morning terminator (Fig.~\ref{fig_1000K}). Compared to our nominal model (Fig.~\ref{fig_chem_CO_N2_HCN}), the HCN abundance is slightly lower and vertically more confined. From this experiment, we conclude that HCN formation -- and night-side photochemistry by extension -- originates from a balance between dynamical and chemical processes that can be tilted one way or another by changing the local temperature. This result also indicates the potential for gauging night-side conditions of ultra-hot Jupiters by measuring their limb abundances via transmission spectroscopy.

\subsection{Limitations of the pseudo-2D assumption}
\label{sec_disc_winds}

Given the pseudo-2D approach used in this work, horizontal advection in our WASP-76~b model is simplified. We have adopted a single value for the horizontal wind speed -- representing the equatorial jet stream -- along the full simulation domain. At high altitudes, however, the possibility exists that magnetic drag slows down the strength of equatorial superrotation \citep{MayKomacekEtal2021ajPhaseCurveWASP-76b}, and the radiative timescale decreases sufficiently so that the jet stream breaks down in favour of thermally direct circulation \citep[e.g.][]{Showman2020_review}. This would yield radial day-to-night wind instead of the uniform eastward wind assumed in our model. Additionally, the wind direction across the morning terminator would be opposite. It is currently unclear at which altitude this circulation flip would occur, since GCM studies do not agree on the strength of these phenomena, and our model of WASP-76~b still exhibits strong superrotation at the top of the simulation domain \citep{SchneiderEtal2022aaNoInflation}. 

Some high-resolution spectroscopy studies have provided evidence for a day-to-night side circulation regime, along with a flip in wind regimes near $\about$mbar pressures \citep{Seidel2021_WASP-76b}. However, it appears that stringent constraints on the wind regime are hampered by degeneracies between the wind speed, altitude-dependence \citep{Gandhi2022_WASP-76b}, and chemical tracer distribution \citep{Wardenier2023_WASP-76b} in the 2D projected disk.

Taking into account the full three-dimensional wind flow of WASP-76~b would yield two main differences with our currently presented model. First, in the case of a circulation regime transition from strong superrotation to direct day-to-night-side flow, the wind direction on the morning terminator would flip such that hot gas crosses both limbs towards the antistellar point on the night side. Such scenario would likely result in the same night-side chemistry as explored in this work, but with a more symmetrical distribution \citep[see ][and their Fig.~18 for a qualitative example]{Tsai2024_2DVULCAN}. Based on our Fig.~\ref{fig_vary_wind}, we argue that a symmetric day-to-night flow would lead to an HCN abundance that is maximal near the antistellar point. Based on our Fig.~\ref{fig_kzz}, HCN would also be further extended in altitude, since the convergence of wind flows near the antistellar point would lead to more vertical mixing. As a consequence, in principle, it may be possible to use HCN, \chem{SO_2}, or other night-side species discussed here, to diagnose the circulation regime of ultra-hot planets based on observed limb asymmetries.

Besides a possible breakdown of equatorial superrotation, a second concern would be that of meridional (i.e.~pole-to-equator) advection via a thermally direct circulation. Naturally, in a two-dimensional model, this cannot be taken into account consistently. In the case of strong meridional circulation, it is possible for the equator to get quenched by gas originating from the relatively cold night-side Rossby gyres \citep[see][for a proof of concept]{Drummond2018_HD189733b}. Given our finding that the horizontal advection timescale and chemical timescale are comparable (Sect.~\ref{sec_jet_speed}), we can speculate that the HCN distribution would be affected by a change in circulation regime. Future work focussing on 3D advection in giant exoplanets \citep[cfr.][]{Steinrueck2021, Zamyatina2023} combined with photochemistry would shed light on the detailed distribution of night-side photochemical species. 

\subsection{Potential impact of ion chemistry}
\label{sec_disc_ions}

Although for this work we have investigated the night-side chemistry of an ultra-hot planet using a neutral photochemical model, it is probable that the atmosphere of WASP-76~b is at least partly ionized. Several ionic species have already been detected in the atmosphere or extended exosphere of WASP-76~b, such as Ca$^+$, Ba$^+$, and Sr$^+$ \citep{Tabernero2021_WASP-76b, AzevedoSilva2022_barium, Kesseli2022_spectralSurvey, Pelletier2023_VO_WASP-76b, Deibert2023_WASP-76b}. As such, potentially important chemical species and reactions involving ions are not taken into account here. We discuss the potential impact of ions on our findings.

\subsubsection{Night-side ionization fraction}

\citet{Koskinen2014} and \citet{Lavvas2014} have studied ionization on the canonical hot Jupiter HD~209458~b. The emerging 1D picture for hot gas giant ionospheres is one roughly split into two layers, namely a high-altitude thermally ionized layer dominated by H$^+$, overlying a deeply extended region with high abundances of Na$^+$ and K$^+$. The latter zone is driven by photo-ionization, and can reach up to 10-100~mbar. For HAT-P-7~b, an ultra-hot planet with an equilibrium temperature similar to WASP-76~b, \citet{Helling2021_HAT-P-7b3_ionization} have shown that thermal ionization is sufficient to drive a deep day-side ionosphere that can extend to pressures up to several 100~mbar. Analogous to the picture in our Fig.~\ref{fig_dissociation}, it is possible that the inclusion of photo-ionization would give rise to even more ionization at depth. Thus, it appears that a deep day-side ionosphere is evident on WASP-76~b.

On the night side, however, \citet{Helling2021_HAT-P-7b3_ionization} found no significant degree of thermal ionization. The ionization fraction -- the ratio of ionized to total gas density -- was found to remain lower than $10^{-10}$ for the whole night side, with some evening-morning asymmetry that is correlated with temperature. The inclusion of photo-ionization is unlikely to alter this result for the shielded night side. Horizontal ion advection from the day side, not included in the model of \citet{Helling2021_HAT-P-7b3_ionization}, may thus be the main factor determining the night-side ionization fraction.

\citet{Koskinen2014} calculated the night-side ionization fraction $f_\textrm{day-night}$ based on a simple balance between the advection rate of ions and electrons, and their recombination rate:
\begin{equation}\label{eq_ionization}
    f_\textrm{day-night} = \frac{n_\textrm{e,day}}{n_\textrm{e,night}}
                        \approx 1 + \frac{r n_\textrm{e,day} \Phi_\textrm{eff}}{u_\phi} \Delta \phi,
\end{equation} where $n_e$ is the electron density, $r$ the radial distance, $u_\phi$ the horizontal wind speed, and $\Delta\phi$ the angular distance over which the gas is advected ($\pi$/2 for the antistellar point). The effective recombination rate $\Phi_\textrm{eff}$ is the mean recombination rate weighted by the day-side ionization fraction of all ionic species.
On a hot Jupiter, \citet{Koskinen2014} found that the night-side ionization fraction is indeed several orders of magnitude lower than the day-side value for pressures between $10^{-1}$-$10^{-6}$~bar. The difference grows smaller at high altitudes, but becomes less than an order of magnitude only in the extended thermosphere ($p < 10^{-8}$~bar).


\subsubsection{Night-side ion chemistry}

In case the night-side ionization fractions of ultra-hot Jupiters do remain high, we may expect the dominant ions to be ionized atoms H$^+$, C$^+$, O$^+$, together with ionized metals, especially the alkali metals Na$^+$ and K$^+$ \citep{GarciaMunoz2007, Koskinen2013, Lavvas2014}. Another prominent ion may be H$_3^+$, which is an efficient radiative coolant in the lower thermosphere \citep{Yelle2004, Koskinen2013, Bourgalais2020}. However, given the high equilibrium temperature of ultra-hot Jupiters, it is possible that the H$_3^+$ abundance will remain low, since it is formed via \ce{H2+ + H2 -> H3+ + H} and most of the hydrogen may be in atomic form due to thermal and photodissociation (Figs.~\ref{fig_chem_H2O_OH_H} and \ref{fig_app_atomic}). On the other hand, radiative cooling by H$_3^+$ may also prevent the onset of a hot thermospheric layer on the night-side, allowing for the survival of molecules and inducing neutral-ion chemistry.

The tendency for night-side ion-neutral chemistry on ultra-hot gas giants may be informed by our understanding of photochemistry in the atmosphere of Titan. While not exactly analogues, a comparison between both types of atmospheres can give insight in the destruction of HCN and the production of daughter species in an ionized environment. The main difference affecting the chemistry in this work is the temperature, which is much higher than on Titan. As a consequence, the reaction rates are changed, and potentially even ill-determined at high temperatures. 
Nevertheless, we briefly explore ionic reactions with HCN, based on Titan \citep{Vuitton2019_Titan}.

If indeed H$_3^+$ exists as a dominant ion in the night-side ionosphere, this may lead to proton exchange with HCN: \ce{H3+ + HCN -> HCNH+ + H2}. If the atmosphere is too hot for H$_3^+$, the dominant formation mechanism for HCNH$^+$ would probably be the direct proton (H$^+$) capture by HCN. The main destruction pathways for the HCNH$^+$ ion are through electron recombination \citep{Vuitton2019_Titan}:
\begin{align}
    \ce{
        HCNH+ + e-  &-> HCN + H     \textrm{,}  \\
                    &-> CN + H2     \textrm{,~or}\\
                    &-> CN + H + H  \textrm{.}  
        }
\end{align}
As such, the HCNH$^+$/HCN fraction will depend strongly on the amount of electrons to protons. On ultra-hot planets, it is likely that a significant fraction of the electrons originate from ionized (alkali) metals in the deeper atmospheric layers \citep{Lavvas2014}. As such, we may expect electron recombination to be favoured over proton capture, and thus HCN over HCNH$^+$.

In fact, the presence of ionized metal vapour and high electron densities can give rise to appreciable fractions of H$^-$ in ultra-hot exoplanets \citep{Arcangeli2018, ParmentierEtal2018aaThermalDissociation}. If this ion can persist on the night side alongside atomic carbon and nitrogen, this may lead to the formation of CN$^-$, which is one of the dominant negatively charged species on Titan \citep{Vuitton2009_Titan_negative_ions}. Transmission spectroscopy observations of WASP-76~b, however, did not provide evidence for a high H$^-$ abundance \citep{EdwardsEtal2020ajARES1WASP-76b, FuEtal2021ajWASP-76b}.


Finally, we stress that it is difficult to conclude about the effect of ion chemistry without a self-consistent radiative-chemical model. Clearly, the presence of each of these ionic species on the night side will strongly depend on the local temperature, their recombination rate, and local electron density. In particular, radiative heating and cooling become important feedback phenomena to incorporate. As such, it remains uncertain whether ions would greatly affect the night-side chemistry of ultra-hot planets presented in this work. But since these objects clearly at least partly experience ionization, we encourage future studies involving ion chemistry on hot planets, and in particular those that incorporate horizontal ion transport.



                          
\section{Conclusions}
\label{sec_conclusion}

In this work, we have modelled the chemical composition of the ultra-hot giant WASP-76~b in a pseudo-2D framework, taking into account photochemistry and day-to-night transport. As such, our model is complementary with 3D climate models of ultra-hot Jupiters that assume equilibrium chemistry. 
Previous work has shown that, generally, products of photochemistry are produced and transported to the planetary night side of warm and hot gaseous planets \citep{BaeyensEtal2022mnrasPhotochemistry, Tsai2023_SO2_theory}. 
Here, we demonstrate that, despite the elevated temperatures of WASP-76~b, disequilibrium chemistry still has a potentially important impact on its chemical inventory. We formulate two main conclusions.

Firstly, we find that photolysis on the day-side of WASP-76~b takes place in tandem with thermal dissociation, causing the majority of the upper day-side gas to be atomic, rather than molecular. Molecular dissociation occurs even for strong, triple-bonded molecules such as CO and \chem{N_2}. Importantly, photochemistry is non-negligible for CO, since we find that thermal dissociation alone is insufficient to break down CO. This finding has potential implications for recent studies that advocate for CO as a stable molecule to break degeneracies in high-resolution spectroscopy \citep{SavelEtal2023apjAsymmetries, Wardenier2023_WASP-76b}. Furthermore, the additional component of \chem{H_2} dissociation provided through photolysis may also impact the rate of heat redistribution between the day and night sides of ultra-hot planets.

Secondly, we find that advection of radicals from the hot day-side hemisphere of WASP-76~b along the equatorial jet stream can kick-start disequilibrium chemistry on the cooler night side. In particular, our model exhibits an asymmetric HCN concentration with a maximum near the morning limb. This mechanism hinges on the presence of photochemistry in order to free up carbon from CO. As such, the scenario is proof of concept for the asymmetric HCN absorption measured on this planet \citep{SanchezLopezEtal2022aaWASP76b}. However, the night-side temperature and overall wind speed appear to have a large effect on the HCN distribution. Thus, a systematic and statistical approach would be required to match the model to the observational data.

Since we show that the night-side hemispheres of ultra-hot Jupiters can be more complex than simple chemical equilibrium models would predict, potential future research avenues would benefit from taking into account additional kinetic processes such as photochemistry and chemical transport, but also molecular recombination, condensation, and ionization. As the technique of high-resolution spectroscopy matures and single planets yield a multitude of chemical species \citep[e.g.][on WASP-76~b]{Landman2021_OH_WASP-76b, SanchezLopezEtal2022aaWASP76b, Kesseli2022_spectralSurvey, Yan2023_CO_WASP-76b_WASP-18b, Pelletier2023_VO_WASP-76b}, it is clear that 2D and 3D models, as well as chemical kinetic schemes involving vanadium, iron, and silicates will become valuable tools to understand the atmospheres of these planets.


\begin{acknowledgements}
      RB acknowledges support from the \textit{Origins} investment incentive.
      The authors acknowledge the use of the Lisa Computing cluster hosted by SURFsara.
      For this work we have made use of the GCM post-processing library \textit{gcm\_toolkit} \citep{SchneiderBaeyensKiefer2022zenodoGcmtoolkit}.
\end{acknowledgements}

%
%

\bibliographystyle{aa} 
\bibliography{my_citations} 


\begin{appendix}


\section{Abundance maps for atomic species}
\label{app_atomic}

\begin{figure*}
    \centering
    \includegraphics[width=\textwidth]{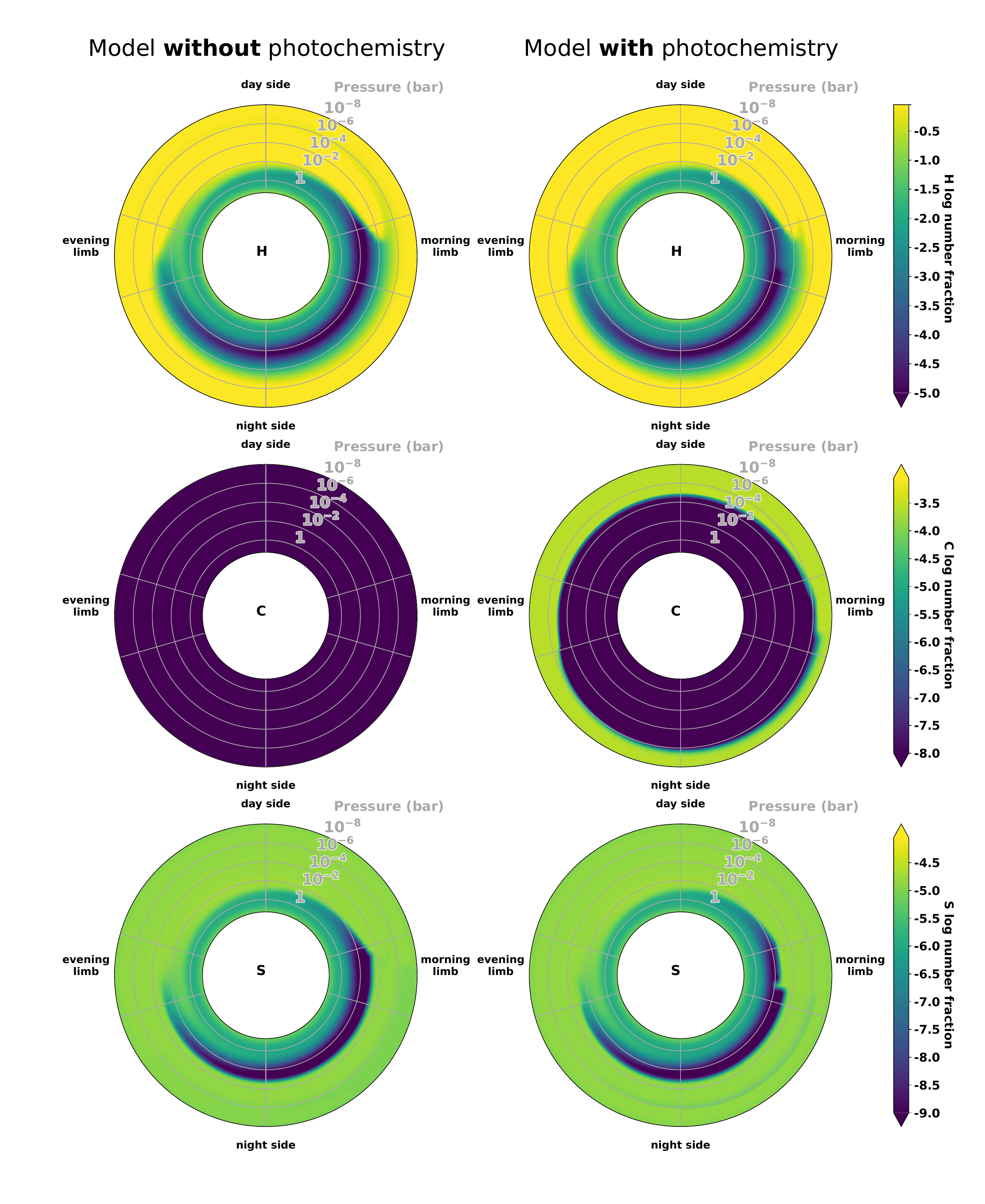}
    \caption{Polar abundance maps of atomic hydrogen, carbon, and sulfur in our WASP-76~b model. We compare models excluding (\textit{left}) and including (\textit{right}) photochemistry. The H and C abundances have been computed using the \citet{Venot2020_network}-network, while the S abundance was computed using the VULCAN chemical network of \citet{Tsai2021}.}
    \label{fig_app_atomic}
\end{figure*}

\begin{figure*}
    \centering
    \includegraphics[width=\textwidth]{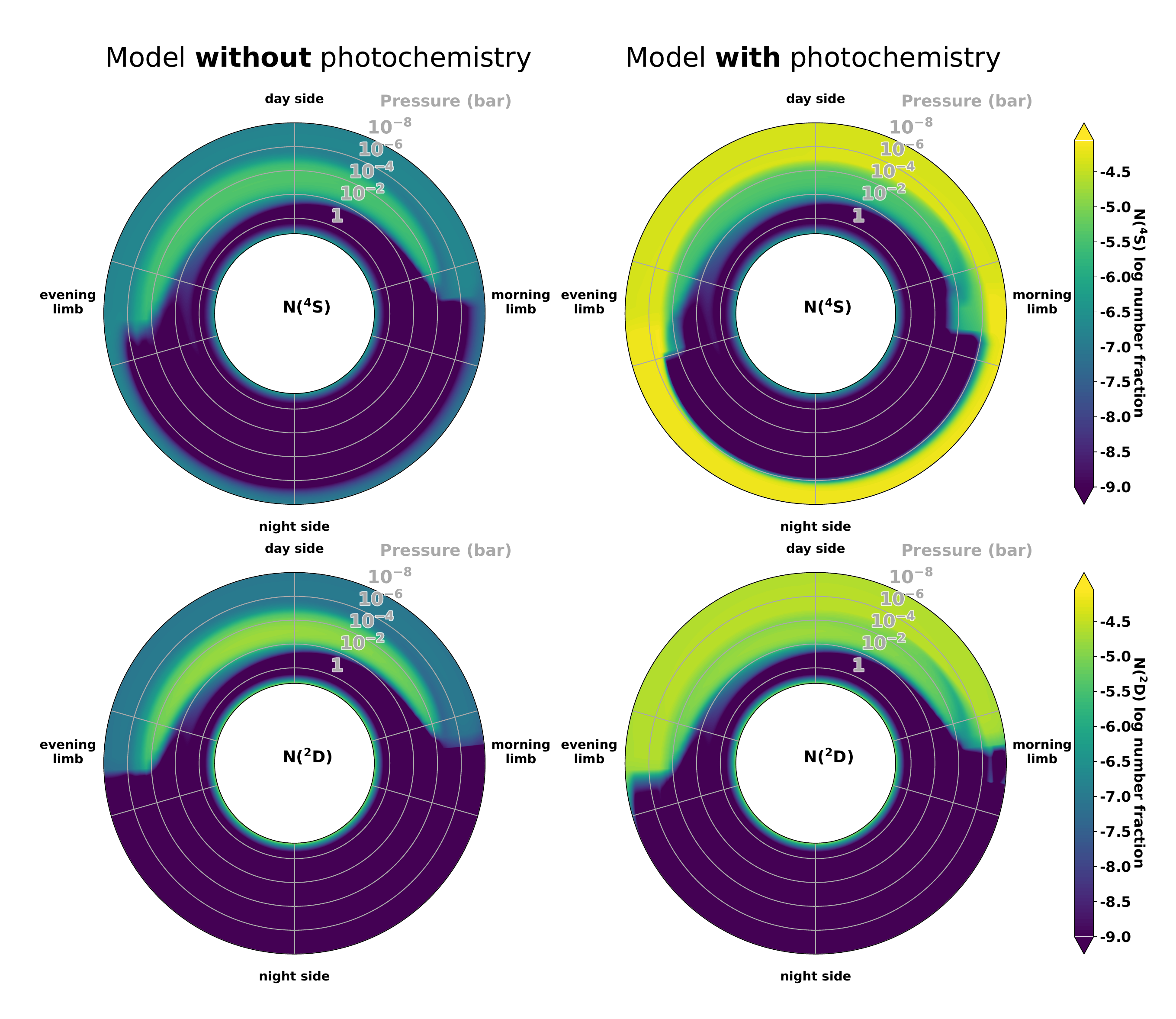}
    \caption{Polar abundance maps of the ground ($^4$S) and excited ($^2$D) states of atomic nitrogen in our WASP-76~b model. We compare models excluding (\textit{left}) and including (\textit{right}) photochemistry.}
    \label{fig_app_nitrogen}
\end{figure*}

\begin{figure*}
    \centering
    \includegraphics[width=\textwidth]{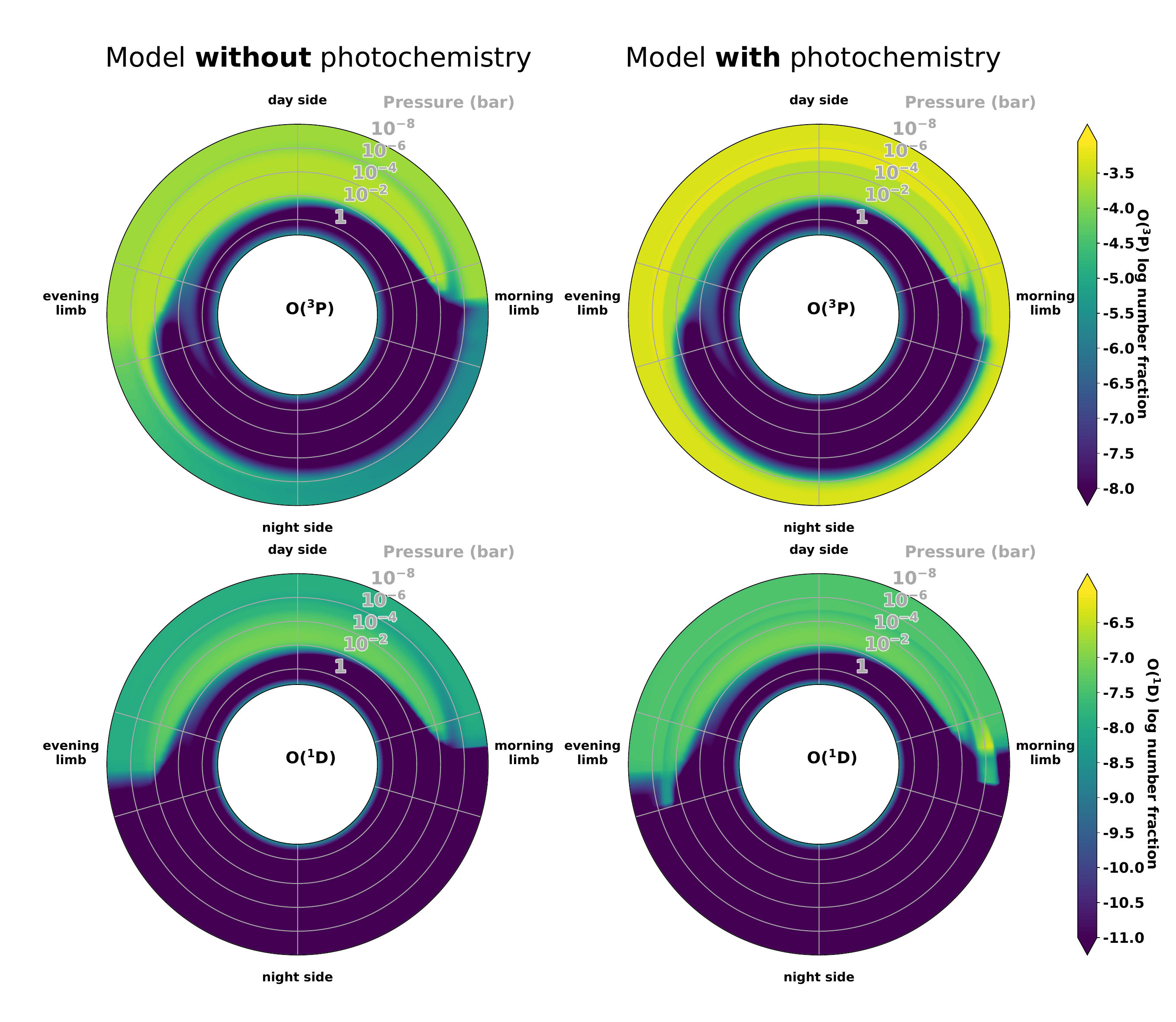}
    \caption{Polar abundance maps of the ground ($^3$P) and excited ($^1$D) states of atomic oxygen in our WASP-76~b model. We compare models excluding (\textit{left}) and including (\textit{right}) photochemistry.}
    \label{fig_app_oxygen}
\end{figure*}

Here, we present additional abundance maps of all atomic species in our chemical models. The atomic species comprise hydrogen, carbon, sulfur (Fig.~\ref{fig_app_atomic}), nitrogen (Fig.~\ref{fig_app_nitrogen}), and oxygen (Fig.~\ref{fig_app_oxygen}). Since the chemical network of \citet{Venot2020_network} distinguishes between ground state and excited state N and O, we plot the abundances of both of them. 

Analyzing the atomic species concentrations in general, it is clear that photodissociation causes a general increase in all atomic species, as it combines with thermal dissociation to break up molecules. Carbon in particular (Fig.~\ref{fig_app_atomic}) is strongly affected by the inclusion of photochemistry, since thermal dissociation alone is insufficient to break up CO. Indeed, the carbon abundance increases by orders of magnitude in the photochemistry case. To a lesser degree, we see the same phenomenon happening for nitrogen (Fig.~\ref{fig_app_nitrogen}).

Upon comparing the N and O ground state and excited state distributions, we find that ground state \chem{N(^4S)} and \chem{O(^3P)} are abundant on both hemispheres of the planet. This signifies an efficient transport from the day to the night side via the equatorial jet stream. On the other hand, their metastable counterparts, \chem{N(^2D)} and \chem{O(^1D)} respectively, are only abundant on the day-side hemisphere, but quickly undergo collisional or radiative de-excitation before they reach the night side.


\section{Vertical mixing sensitivity}
\label{app_kzz}

\begin{figure}
    \centering
    \includegraphics[width=\columnwidth]{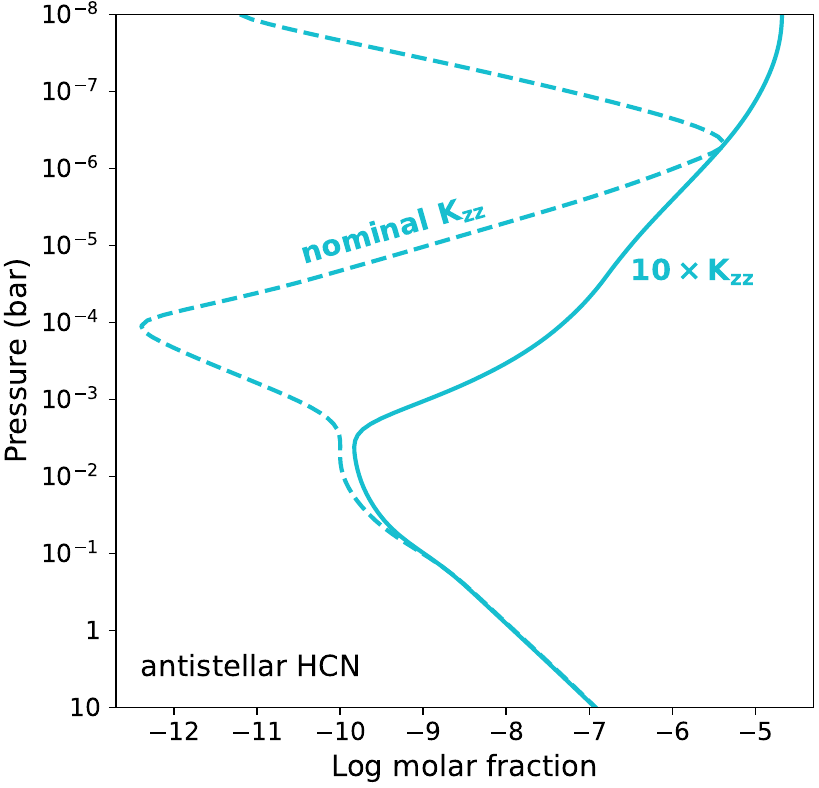}
    \caption{The HCN abundance profile at the antistellar point of the nominal pseudo-2D model (\textit{dashed}) and a model with $\kzz$ that is ten times higher (\textit{solid}). Both models attain the same HCN abundance near $\mu$bar level, but the model with increased vertical mixing has a higher integrated abundance overall.}
    \label{fig_kzz}
\end{figure}

In this Appendix, we present a sensitivity test for the vertical mixing efficiency $\kzz$ in our model, and its effect on the night-side HCN abundance. We have run an additional pseudo-2D chemistry model, set up with a $\kzz$-profile that is ten times higher. Thus, the original parametrization \citep{ParmentierEtal2013aa3DMixing} is now given by $\kzz = 5 \cdot 10^9 \left( P_\textrm{bar} \right)^{-0.5}$~cm$^2$s$^{-1}$, with $P_\textrm{bar}$ the pressure in bar and an imposed maximum of $10^{12}$~cm$^2$s$^{-1}$.

The HCN abundance at the antistellar point is seen in Fig.~\ref{fig_kzz}. Near $\mu$bar pressures, which is where HCN is produced most efficiently, both models attain a similar molar fraction of $10^{-5.5}$. The nominal model has a distinct peaked profile, since zonal advection dominates over vertical mixing, and there is limited time for HCN to diffuse vertically during the night side crossing. The model with stronger vertical mixing, however, is more efficient in distributing HCN vertically, and as a result has overall higher HCN abundances.

The vertical mixing timescales, estimated as $\timescale{vert} \approx H^2/\kzz$ with $H$ the local scale height, are about 10$^5$~s and 10$^4$~s for the models with nominal and strong vertical mixing respectively. Thus, within an order of magnitude, the vertical mixing timescale is of similar magnitude as the zonal advection timescale (Sect.~\ref{sec_jet_speed}) in the nominal model, explaining the HCN distribution that is confined to the upper atmosphere. The strong vertical mixing case, on the other hand, has a vertical mixing timescale that is markedly shorter than the zonal advection timescale, resulting in an abundant HCN profile that is more extended in pressure (Fig.~\ref{fig_kzz}).


\end{appendix}

\end{document}